\documentstyle[12pt]{article}

\newcommand{\sect}[1]{\setcounter{equation}{0}\section{#1}}

\newcommand{\EQ}{\begin{equation}}
\newcommand{\EN}{\end{equation}}
\newcommand{\bea}{\begin{eqnarray}}
\newcommand{\ena}{\end{eqnarray}}
\newcommand{\vs}[1]{\vspace{#1 mm}}
\newcommand{\hs}[1]{\hspace{#1 mm}}

\renewcommand{\b}{\beta}

\newcommand{\e}{\epsilon}
\def\bbox{{\,\lower0.9pt\vbox{\hrule \hbox{\vrule height 0.2 cm
\hskip 0.2 cm \vrule height 0.2 cm}\hrule}\,}}
\newcommand{\dsl}{\pa \kern-0.5em /}

\newcommand{\shalf}{\frac{1}{2}}
\newcommand{\pa}{\partial}
\renewcommand{\t}{\theta}

\newcommand{\nn}{\nonumber\\}
\newcommand{\tH}{\tilde H}

\begin{document}

\topmargin 0pt
\oddsidemargin 5mm

\begin{titlepage}

\setcounter{page}{0}
\begin{flushright}
OU-HET 267 \\
hep-th/9706153
\end{flushright}

\vs{10}
\begin{center}
{\Large
Towards the Classification of Non-Marginal Bound States of M-branes and
Their Construction Rules}
\vs{15}

{\large
Nobuyoshi Ohta\footnote{e-mail address: ohta@phys.wani.osaka-u.ac.jp}
and Jian-Ge Zhou\footnote{e-mail address:
 jgzhou@phys.wani.osaka-u.ac.jp} \footnote{JSPS postdoctral fellow}} \\
\vs{10}
{\em Department of Physics, Osaka University, \\
Toyonaka, Osaka 560, Japan}

\end{center}
\vs{15}
\centerline{{\bf{Abstract}}}

We present a systematic analysis of possible bound states of M-brane solutions
(including waves and monopoles) by using the solution generating technique
of reduction of M-brane to 10 dimensions, use of T-duality and then lifting
back to 11 dimensions. We summarize a list of bound states for
one- and two-charge cases including tilted brane solutions.
Construction rules for these non-marginal solutions are also discussed.

\end{titlepage}
\newpage

\sect{Introduction}

There has been a great advance in our understanding of possible classical
solutions of superstrings and supergravities. These classical solutions play
important role in strong coupling dynamics of string theories~\cite{HT,W}.

It is now believed that the best candidate for a unified theory underlying
all physical phenomena is no longer 10-dimensional string theory
but rather 11-dimensional M-theory whose low energy limit is given by
the 11-dimensional supergravity. Thus it is expected that these classical
solutions can be understood most easily in the 11-dimensional supergravity.
In fact, we will see that various different 10-dimensional solutions can be
obtained from an 11-dimensional solution. It is thus simpler to consider
11-dimensional solutions, on which we focus in this paper.

It has been known that this theory admits various $p$-brane
solutions~\cite{DS,GU,TO,DKL}, collectively referred to as M-branes. These
solutions have been shown to be understood as the intersections of the
fundamental $2_M$ and $5_M$ brane solutions~\cite{PT,TS1,GKT}.

Systematic (intersection) rules are given for making various marginal
solutions as intersecting M-branes in 11
dimensions~\cite{TS1,GKT,AEH,O,BR,CT,AR,OS}. However, it is also known that
there are a large class of non-marginal solutions typically characterized by
the mass formula $m \sim \sqrt{Q_e^2 + Q_m^2}$ in terms of the electric
$Q_e$ and magnetic $Q_m$ charges. The rules for constructing these solutions
are not explicitly given. It would be quite interesting to try to formulate
such rules. For this purpose, it is important to first understand various
solutions of this kind.

{}For 10-dimensional solutions, we can use T-duality~\cite{BHO} to generate
new solutions from known ones~\cite{BJO,CH,TS2}. In particular, some
non-marginal solutions have been explicitly constructed in this
way~\cite{RT,TS3}, but a unified understanding still seems to be lacking.
{}For the 11-dimensional solutions, there is no analogous technique, but we
can make reductions to 10 dimensions and then make various duality
transformations to get new 11-dimensional solutions~\cite{BHO}. Though several
solutions of this kind have been found, they are scattered in various
literature~\cite{RT,IT,C,CP,TS3,BC,CC,GG,BD,BL,BLL,BM} and no systematic
classification has been attempted.

The purpose of this paper is to present a rather systematic analysis of
possible ``bound states'', which are typical non-marginal solutions in
11-dimensional supergravity. Some of the solutions
have been derived by using sophisticated symmetry in lower
dimensions~\cite{CC}. We will see that all the bound states of this kind
can be obtained step by step by using simple duality rules in 10 dimensions
without using larger symmetry realized in lower dimensions.
We also present many new solutions and discuss the methods how to make
these non-marginal solutions, which we call construction rules.

We first summarize fundamental solutions in 11-dimensional supergravity which
will be our starting point.

$2_M$-brane solution:
\bea
2_M: \hs{5}
ds_{11}^2  &=& H^{1/3} \left[ H^{-1} \left(- dt^2 + dy_1^2 + dy_2^2 \right)
 + \sum_{i=1}^8 dx_i^2 \right], \nn
C &=& \frac{1-H}{H} dt \wedge dy_1 \wedge dy_2,
\label{2M}
\ena
where $H$ is a harmonic function depending on the transverse coordinates
$x_1, \cdots, x_8$.

$5_M$-brane solution:
\bea
5_M: \hs{5}
ds_{11}^2 &=& H^{2/3} \left[ H^{-1}\left(- dt^2 + dy_1^2 + \cdots +
dy_5^2 \right) + \sum_{i=1}^5 dx_i^2 \right], \nn
dC &=& *dH,
\label{5M}
\ena
where the dual is taken with respect to the transverse coordinates
$x_1, \cdots, x_5$.

Wave solution:
\bea
(0_w):\hs{5}
ds_{11}^2 &=& -dt^2 + dy_1^2 + (H-1)(dt-dy_1)^2 + \sum_{i=1}^9 dx_i^2, \nn
dC &=& 0.
\label{w}
\ena

Monopole solution:
\bea
(0_m):\hs{5}
ds_{11}^2 &=& -dt^2 + \sum_{n=1}^6 dy_n^2 + H^{-1}(dz + A_i dx_i)^2
 + H \sum_{i=1}^3 dx_i^2, \nn
{}F_{ij} &\equiv& \pa_i A_j - \pa_j A_i = \e_{ijk}\pa_k H.
\label{m}
\ena

Using the rules summarized in the appendix~\cite{BHO}, we can make
reduction of (\ref{2M}) in the direction of world-volume, say $y_2$ to
get a fundamental string $1_F$ in 10 dimensions; if we do this in transverse
direction $x_1$, we obtain a D2-brane $2_D$. Let us write this as~\cite{CP}
\bea
2_M(y_1,y_2) \left\{
\begin{array}{cl}
\stackrel{y_2}{\to} & 1_F(y_1) \\
\stackrel{x_1}{\to} & 2_D (y_1,y_2) \hs{2},
\end{array} \right.
\label{re2M}
\ena
where we have explicitly written the world-volume coordinates.
It is useful to keep track of these coordinates to examine what solutions
are possible. Similarly we can derive the rules for other solutions:
\bea
5_M(y_1,\cdots,y_5) \left\{
\begin{array}{cl}
\stackrel{y_5}{\to} & 4_D(y_1, \cdots, y_4) \\
\stackrel{x_1}{\to} & 5_S (y_1,\cdots, y_5) \hs{2},
\end{array} \right.
\label{r5M}
\ena
\bea
0_w(y_1) \left\{
\begin{array}{cl}
\stackrel{y_1}{\to} & 0_D \\
\stackrel{x_1}{\to} & 0_w (y_1) \hs{2},
\end{array} \right.
\label{rw}
\ena
\bea
0_m(y_1,\cdots,y_6,z) \left\{
\begin{array}{cl}
\stackrel{y_6}{\to} & 0_m(y_1,\cdots,y_5,z) \\
\stackrel{z}{\to} & 6_D (y_1,\cdots,y_6) \hs{2},
\end{array} \right.
\label{rem}
\ena
where $5_S$ is a solitonic 5-brane solution and other solutions are similar
ones in type IIA string.
These rules can also be used to read off what 11-dimensional solutions are
obtained from 10-dimensional ones. Note that the rules give one-to-one
correspondence between the solutions in 10 and 11 dimensions.

{}For T-duality, we can also consider them in various directions. For the
world-volume direction, the procedure is clear. For the transverse (space-time)
direction, we ``delocalize'' the solution, i. e. we include the coordinate
in the direction of the isometry and suppose that the harmonic functions
involved in the solution do not depend on the coordinate, and then use the
ordinary duality rules. Using the rules in the appendix, we find~\cite{CP}
\bea
1_F(y_1) \left\{
\begin{array}{cl}
\stackrel{y_1}{\to} & 0_w(y_1) \\
\stackrel{x_1}{\to} & 1_F (y_1) \hs{2},
\end{array} \right.
\ena
\bea
5_S(y_1,\cdots,y_5) \left\{
\begin{array}{cl}
\stackrel{y_5}{\to} & 5_S(y_1, \cdots, y_5) \\
\stackrel{z}{\to} & 0_m (y_1,\cdots, y_5,z) \hs{2},
\end{array} \right.
\ena
\bea
0_w(y_1) \left\{
\begin{array}{cl}
\stackrel{y_1}{\to} & 1_F (y_1) \\
\stackrel{x_1}{\to} & 0_w (y_1) \hs{2},
\end{array} \right.
\ena
\bea
0_m(y_1,\cdots,y_5,z) \left\{
\begin{array}{cl}
\stackrel{y_5}{\to} & 0_m(y_1,\cdots,y_5,z) \\
\stackrel{z}{\to} & 5_S (y_1,\cdots,y_5) \hs{2},
\end{array} \right.
\ena
\bea
p_D(y_1, \cdots, y_p) \left\{
\begin{array}{cl}
\stackrel{y_p}{\to} & (p-1)_D(y_1, \cdots,y_{p-1}) \\
\stackrel{x_1}{\to} & (p+1)_D(y_1,\cdots,y_p,x_1) \hs{2}.
\end{array} \right.
\ena

If we make reductions and T-dualities in mixed directions characterized by
an angle, we can get non-marginal bound states of the above solutions.

Thus our procedure to produce new solutions in 11 dimensions are just to
(1) make reduction, (2) make T-dualities twice in various directions and
then (3) lift it back to 11 dimensions. If we start from various marginal
solutions and continue this process, this is enough to derive all possible
non-marginal solutions. This is the approach advocated by Tseytlin~\cite{TS2},
and we intend to elaborate on this approach in more detail including
non-marginal solutions. It is expected that this will lead to rules how to
construct solutions in 11 dimensions.

We will see in the next section that all the fundamental
solutions~(\ref{2M})--(\ref{m}) are connected
by duality transformation, and hence starting from just any single solution
one can reproduce all other solutions. These solutions are related with
more general bound states with angles which characterize how
the solutions are interpolated. These solutions are non-marginal and have
1/2 supersymmetry. We will refer to these solutions obtained from the
fundamental solutions listed in (\ref{2M})--(\ref{m}) as non-marginal
solutions with single charge and summarize them in sect.~2.

If we start from two-charge solutions like $2_M\perp 2_M$, we get various
bound states with 1/4 supersymmetry. We will derive possible non-marginal
bound states obtained from marginal solutions of this kind in sect.~3.
{}Further generalization including boost is discussed in sect.~4.

In sect.~5, we discuss the construction rules for the non-marginal solutions.
{}Finally sect.~6 is devoted to discussions. In particular, we examine
the effect of S-duality and the ADM mass for the non-marginal solutions.

\sect{Non-marginal solutions with single charge}

In this section, we present solutions characterized by one independent
charge and hence with 1/2 supersymmetry. All the solutions are connected
by T-duality and can be obtained from $2_M$-brane solution (or any one of the
fundamental solutions).

Let us first discuss what kind of bound states are possible if
we start from marginal solutions (\ref{2M})--(\ref{m}) with single charge.
Suppose we start from $2_M$ solution~(\ref{2M}). In the first step of
reduction, we get from (\ref{re2M}) a bound state of $(1_F+2_D)_A$, where
subscript $A$ indicates that it is type IIA solution. If we make T-duality
in all possible directions, we find a bound state of $(0_w+1_F+1_D+3_D)_B$,
which transforms into $(1_F+0_w+0_D+2_D+4_D)_A$ bound state in the second
T-duality transformation. This in turn can be lifted to 11-dimensional
non-marginal solution of two $0_w$, seven $2_M$ and one $5_M$. Instead of
lifting there, we can further make third T-duality to
get more general bound state of two $0_w$, sixteen $2_M$, fifteen $5_M$
and one monopole. In deriving this result, it is very useful to keep track
of world-volume coordinates at each steps.

It is clear from this example that all the fundamental solutions
(\ref{2M}) -- (\ref{m}) are related by the T-duality transformations.
We can repeat this procedure to get other possible bound states starting
from other fundamental solutions listed in (\ref{5M})--(\ref{m}). Below we
discuss all the two-body bound states and some new three-body bound state
solutions in order to show explicitly how the above procedure actually works.

{}For this purpose, we first delocalize the solution (\ref{2M}) (include
$x_1$ in the isometry direction) and rotate the coordinates $y_2$ and $x_1$:
\bea
ds_{11}^2 &=& H^{1/3} \left[ H^{-1} \left(- dt^2 + dy_1^2
 + (dy_2 \cos\t + dx_1 \sin\t )^2 \right) \right. \nn
&& \left. \hs{10} + (-dy_2 \sin\t + dx_1 \cos\t )^2
 + \sum_{i=2}^8 dx_i^2 \right], \nn
C &=& \frac{1-H}{H} dt \wedge dy_1 \wedge (dy_2 \cos\t + dx_1 \sin\t) .
\label{r2M}
\ena
We note that, though trivial as 11-dimensional solution, this may be
regarded as a bound state of $2_M+2_M$ lying in the directions of
$y_1-y_2$ for $\t=0$ and $y_1-x_1$ for $\t=\frac{\pi}{2}$.

Reduction in the $y_2$ direction yields a nontrivial bound state
$(1_F + 2_D)_A$ in type IIA string:
\bea
ds_{A}^2 &=& \tH_\t^{1/2} \left[ H^{-1} (- dt^2 + dy_1^2)
 + \tH_\t^{-1} dx_1^2 + \sum_{i=2}^8 dx_i^2 \right], \nn
C &=& \frac{1-H}{H} dt \wedge dy_1 \wedge dx_1 \sin\t , \hs{5}
B^{(1)} = \frac{1-H}{H} dt \wedge dy_1 \cos\t , \nn
A^{(1)} &=& \tH_\t^{-1} (1-H) dx_1 \sin\t \cos\t , \hs{5}
e^{2\phi} = H^{-1}\tH_\t^{3/2},
\label{12}
\ena
where
\bea
{\tH_\t} = \cos^2 \t + H \sin^2 \t.
\label{ht}
\ena
{}For $\t=0$, this is the fundamental string lying in $y_1$;
for $\t=\frac{\pi}{2}$, this becomes D2-brane lying in $y_1-x_1$.

\subsection{Bound states of $0_w$ with $2_M$, $5_M$ and monopole $(0_m)$}

These solutions can be obtained by the following sequence of reduction,
T-duality and lifting:
\bea
2_M \stackrel{R_\t}{\to} (1_F + 2_D)_A \stackrel{T_{y_1}}{\to}
(0_w + 1_D)_B \stackrel{T_{x_2}}{\to} (0_w + 2_D)_A
 \stackrel{{\rm lift}(y_2)}{\to} 0_w + 2_M ,
\label{02}
\ena
where the notation is as follows: $R_\t$ stands for reduction in (\ref{12}),
$T_{y_1}$ for a T-duality along $y_1$ and finally lift$(y_2)$ for lifting
the result to 11 dimensions by adding the coordinate $y_2$.

The resulting solution is the bound state $0_w + 2_M$:
\bea
ds_{11}^2 &=& {\tH_\t}^{1/3} \left[ \tH_\t^{-1} \left( -dt^2
 + dy_1^2 + (H-1)(dt \cos \t + dy_1)^2 +dx_1^2 + dx_2^2 \right) \right.\nn
&& \left. \hs{10}+ dy_2^2 + \sum_{i=3}^8 dx_i^2 \right], \nn
C &=& \frac{(1-H)\sin\t}{\tH_\t} (dt + dy_1 \cos \t)
\wedge dx_2 \wedge dx_1 .
\label{s02}
\ena
This solution was discussed in ref.~\cite{RT}, but it was derived in
a different route. Thus there are many routes to give the same solutions.

Instead of lifting, we can further make T-duality to get
\bea
(0_w + 2_D)_A \stackrel{T_{x_3}}{\to} (0_w + 3_D)_B \stackrel{T_{x_4}}{\to}
(0_w + 4_D)_A \stackrel{{\rm lift}(y_2)}{\to} 0_w + 5_M .
\label{05}
\ena
The solution now takes the form
\bea
ds_{11}^2 &=& {\tH_\t}^{2/3} \left[ \tH_\t^{-1} \left( -dt^2
 + dy_1^2 + (H-1)(dt \cos \t + dy_1)^2 + dy_2^2 
 + \sum_{i=1}^4 dx_i^2 \right) \right. \nn
&& \hs{15} \left. + \sum_{i=5}^8 dx_i^2 \right], \nn
dC &=& * d H \wedge (dy_1 + dt \cos \t) \sin\t ,
\label{s05}
\ena
again in agreement with ref.~\cite{RT}. Here $H$ and $\tH_\t$ depend only
on the transverse space $x_5, \cdots, x_8$ and $*$ is a dual in that space.

Again instead of lifting, we can further make T-duality to get the new
solution of bound state of wave and monopole:
\bea
(0_w + 4_D)_A \stackrel{T_{x_5}}{\to} (0_w + 5_D)_B \stackrel{T_{x_6}}{\to}
(0_w + 6_D)_A \stackrel{{\rm lift}(z)}{\to} 0_w + 0_m .
\label{wm}
\ena
The metric is given by
\bea
ds_{11}^2 &=& -dt^2 + dy_1^2 + (H-1)(dt \cos \t + dy_1)^2
 + \sum_{i=1}^6 dx_i^2 + \tH_\t \sum_{i=7}^8 dx_i^2 \nn
&& + \tH_\t^{-1} ( dz + A_{y_1} \sin\t dy_1 + A_t \sin\t \cos\t dt)^2,
\label{swm}
\ena
where the gauge fields depend only on $x_7$ and $x_8$, and satisfy
\bea
&& \pa_{x_8} A_{y_1} = - \pa_{x_7} H, \hs{5}
\pa_{x_7} A_{y_1} =  \pa_{x_8} H, \hs{5} \nn
&& \pa_{x_8} A_{t} = \pa_{x_7} H, \hs{5}
\pa_{x_7} A_{t} = - \pa_{x_8} H,
\ena
which describe a monopole solution in a special gauge.

\subsection{Bound states of $0_m$ with $2_M$, $5_M$ and wave $(0_w)$}

In order to derive bound states of monopole and others, we can continue
the duality transformation to the above solutions, but
it is easier to start from monopole solution itself.

We rotate $y_6$ and $z$ in the monopole solution (\ref{m}):
\bea
ds_{11}^2 &=& -dt^2 + \sum_{n=1}^5 dy_n^2 + (dy_6 \sin\t + dz \cos\t)^2 \nn
&& \hs{5} + H^{-1}(-dy_6 \cos\t + dz \sin\t + A_i dx_i)^2
 + H \sum_{i=1}^3 dx_i^2.
\ena
Again this is a trivial bound state of $0_m+0_m$.

Reduction in the direction $y_6$ yields the type IIA solution $0_m+6_D$:
\bea
ds_{A}^2 &=& (H \tH_\t)^{1/2} \left[ H^{-1}\left(-dt^2 + \sum_{n=1}^5 dy_n^2
 \right) + (H \tH_\t)^{-1}(dz + A_i \sin\t dx_i)^2
 + \sum_{i=1}^3 dx_i^2 \right], \nn
{}F_{ij} &\equiv& \pa_i A_j - \pa_j A_i = \e_{ijk}\pa_k H, \nn
A^{(1)} &=& (1-\tH_\t^{-1}) \cot\t dz - \tH_\t^{-1}\cos\t A_i dx_i,\hs{3}
e^{2\phi} = H^{-3/2}\tH_\t^{3/2}.
\label{rm}
\ena

To this solution~(\ref{rm}), we now make the the following sequence of
reduction, T-duality and lifting:
\bea
(0_m + 6_D)_A \stackrel{T_{y_5}}{\to}
 (0_m + 5_D)_B \stackrel{T_{y_4}}{\to} (0_m + 4_D)_A
 \stackrel{{\rm lift}(y_0)}{\to} 0_m + 5_M .
\label{m5}
\ena
We find the solution
\bea
ds_{11}^2 &=& H^{2/3} {\tH_\t}^{1/3} \left[ H^{-1} \left\{
 -dt^2 + \sum_{n=0}^3 dy_n^2 \right\} + \tH_\t^{-1}(dy_4^2+dy_5^2) \right. \nn
&& \left.\hs{10} + (\tH_\t H)^{-1}(dz + A_i \sin\t dx_i)^2
 + \sum_{i=1}^3 dx_i^2 \right],\nn
C &=& (\tH_\t^{-1}-1) dy_4 \wedge dy_5 \wedge dz \cot\t
 + A_i \tH_\t^{-1} dy_4 \wedge dy_5 \wedge dx_i  \cos\t,
\label{sm5}
\ena
in agreement with ref.~\cite{CP}.

Instead of lifting, we can further make T-duality to get $0_m +2_M$:
\bea
(0_m + 4_D)_A \stackrel{T_{y_3}}{\to} (0_m + 3_D)_B \stackrel{T_{y_2}}{\to}
(0_m + 2_D)_A \stackrel{{\rm lift}(y_6)}{\to} 0_m + 2_M .
\label{m2}
\ena
The resulting solution is
\bea
ds_{11}^2 &=& H^{1/3}{\tH_\t}^{2/3} \left[ H^{-1} (
 -dt^2 + dy_1^2 ) + \tH_\t^{-1} \sum_{n=2}^6dy_n^2
 + (H\tH_\t)^{-1}(dz + A_i \sin\t dx_i)^2 \right. \nn
&& \left. \hs{10} + \sum_{i=1}^3 dx_i^2 \right],\nn
dC &=& dt \wedge dy_1 \wedge d(H^{-1} A_i dx_i) \sin\t \cos\t
 + dH^{-1} \wedge dt \wedge dy_1 \wedge dz \cos\t.
\label{sm2}
\ena
This is also given in ref.~\cite{CP}.

Again we further make T-duality to get the bound state of wave and monopole:
\bea
(0_m + 2_D)_A \stackrel{T}{\to} (0_m + 1_D)_B \stackrel{T}{\to}
(0_m + 0_w)_A \stackrel{\rm lift}{\to} 0_m + 0_w .
\ena
The metric is given in (\ref{swm}), showing the consistency of the result.

\subsection{$2_M$ and $5_M$ bound states}

The only remaining two-body bound state is that of $2_M$ and $5_M$~\cite{IT}
which can be obtained by
\bea
2_M \stackrel{R(\t=\frac{\pi}{2})}{\to} (2_D)_A \stackrel{T_{x_2}}{\to}
(3_D)_B \stackrel{T_{x_3-x_4}}{\to} (2_D + 4_D)_A
 \stackrel{{\rm lift}(y_2)}{\to} 2_M + 5_M .
\label{25}
\ena
The solution is
\bea
ds_{11}^2 &=& (H{\tH_\t})^{1/3} \left[ H^{-1} (
 -dt^2 + dx_2^2 + dx_3^2) + \tH_\t^{-1} (dx_1^2 + dy_1^2
 + dy_2^2 ) + \sum_{i=4}^8 dx_i^2 \right],\nn
dC &=& dH^{-1} \wedge dx_3 \wedge dx_2 \wedge dt \cos\t - * dH \sin\t \nn
&& \hs{10} + d \tH_\t^{-1} \wedge dx_1 \wedge dy_1 \wedge dy_2 \cot\t.
\label{s25}
\ena
in agreement with ref.~\cite{IT}.

This exhausts all two-body bound states. One may consider bound states
such as $2_M+2_M$, but they are trivial
solutions in 11-dimensions in the same sense of (\ref{r2M}).

We note that it is quite involved to check that (\ref{s25}) is really a
solution to the field equations of 11-dimensional supergravity, and in
particular one has to take into account the Chern-Simon term. Compared with
that approach, the method of T-duality is much simpler.

\subsection{Three-body bound state: $0_w+2_M+5_M$}

The procedure in the previous subsections can be generalized to include
more bound states. For example, let us consider
\bea
&& 2_M \stackrel{R(\t_1)}{\to} (1_F + 2_D)_A \stackrel{T_{y_1}}{\to}
(0_w + 1_D)_B \stackrel{T_{x_2}}{\to} (0_w + 2_D)_A
\stackrel{T_{x_2-x_3}}{\to} (0_w + 1_D + 3_D)_B \nn
&& \hs{5} \stackrel{T_{x_4}}{\to}
(0_w + 2_D + 4_D)_A \stackrel{{\rm lift}(y_2)}{\to} (0_w + 2_M + 5_M).
\label{w25}
\ena
By this procedure we get a new bound state solution $0_w+2_M+5_M$:
\bea
ds_{11}^2 &=& (\tH_{\t_1}\tH_{\t_{12}})^{1/3} \left[ \tH_{\t_1}^{-1}
 \left( -dt^2 + dy_1^2 + (H-1)(dt\cos\t_1 + dy_1)^2 + dx_1^2 + dx_4^2
 \right) \right. \nn
&& \left. + \tH_{\t_{12}}^{-1} (dx_2^2 + dx_3^2 + dy_2^2)
 + \sum_{i=5}^8 dx_i^2 \right],\nn
dC &=& d \left(\frac{1-H}{\tH_{\t_1}}\right) \wedge dx_4 \wedge dx_1
 \wedge ( dt + dy_1 \cos\t_1 ) \sin\t_1\cos\t_2 \nn
&& \hs{10} + *d H \wedge (dy_1 + dt \cos\t_1) \sin\t_1\sin\t_2 \nn
&& \hs{10} +  d \left(\frac{1-\tH_{\t_1}}{\tH_{\t_{12}}}\right)
 \wedge dx_2 \wedge dx_3 \wedge dx_4 \sin\t_2\cos\t_2,
\label{sw25}
\ena
where $\tH_{\t_1}$ is defined as in (\ref{ht}) and
\bea
\tH_{\t_{12}} = \cos^2 \t_2 + \tH_{\t_1} \sin^2 \t_2.
\label{ht12}
\ena
Note that the definition of this harmonic function is similar to $\tH_\t$
in (\ref{ht}).

If we put $\t_1=0$, this reduces to wave solution; $\t_1=\frac{\pi}{2}$
to $2_M+5_M$ solution in (\ref{s25}). Notice also that one can introduce
wave to the $2_M+5_M$ solution (\ref{s25}) in the null isometry direction
by the method of ref.~\cite{GAR}, and that it gives a similar solution to
(\ref{sw25}) but is actually different. The solution $2_M+5_M$ with wave
is given below in (\ref{s2050}).

\subsection{Three-body bound state: $0_m+2_M+5_M$}

Another example of more complicated bound state is obtained by
\bea
&& 0_m \stackrel{R(\t_1)}{\to} (0_m + 6_D)_A \stackrel{T_{y_6}}{\to}
(0_m + 5_D)_B \stackrel{T_{y_4-y_5}}{\to} (0_m + 4_D + 6_D)_A
\stackrel{T_{y_3}}{\to} (0_m + 3_D + 5_D)_B \nn
&& \hs{5} \stackrel{T_{y_2}}{\to}
(0_m + 2_D + 4_D)_A \stackrel{{\rm lift}(y_6)}{\to} (0_m + 2_M + 5_M).
\label{m25}
\ena
The solution takes the form
\bea
ds_{11}^2 &=& (H \tH_{\t_1}{\hat H}_{\t_{12}})^{1/3} \left[ H^{-1}
 ( -dt^2 + dy_1^2) + \tH_{\t_1}^{-1} (dy_2^2 + dy_3^2)
 + {\hat H}_{\t_{12}}^{-1} (dy_4^2 + dy_5^2 + dy_6^2) \right. \nn
&& \left. + (H \tH_{\t_1})^{-1} (dz + A_i \sin\t_1 dx_i)^2
 + \sum_{i=1}^3 dx_i^2 \right], \nn
dC &=& \cos\t_1\cos\t_2 dt \wedge dy_1\wedge dz \wedge dH^{-1}
 - \cos\t_1\sin\t_1 \cos\t_2 dt \wedge dy_1\wedge d(H^{-1}A_i dx_i) \nn
&& + \cos\t_1\sin\t_2 dy_2 \wedge dy_3 \wedge d(\tH_{\t_1}^{-1}A_i dx_i)
 - \cot\t_1\sin\t_2 dy_2 \wedge dy_3\wedge dz \wedge d\tH_{\t_1}^{-1} \nn
&& + \sin\t_2 \cos\t_2 d\left(\frac{\tH_{\t_1}-H}{\tH_{\t_{12}}} \right)
 \wedge dy_4\wedge dy_5 \wedge dy_6,
\label{sm25}
\ena
where $\tH_{\t_1}$ is defined as (\ref{ht}) and
\bea
{\hat H}_{\t_{12}} = \tH_{\t_1} \cos^2 \t_2 + H \sin^2 \t_2.
\ena

Putting $\t_1=0$ gives the $2_M+5_M$ solution in (\ref{s25}) while
$\t_1=\frac{\pi}{2}$ monopole solution. It is also possible to
produce a more general bound state describing bound states of all
combinations of $2_M$, $5_M$ and $0_m$ if one includes an additional angle.

\subsection{Is boost necessary?}

Type IIB bound state of $(1_F+1_D)_B$~\cite{SCH} is derived using
boost~\cite{RT}, but this can be obtained from (\ref{12}) by T-duality
in $x_1$ direction. The result is
\bea
ds_{B}^2 &=& \tH_{\t}^{1/2} \left[ H^{-1} ( -dt^2 + dy_1^2)
 + \sum_{i=1}^8 dx_i^2 \right],\nn
\varphi &=& \log H^{-1/2}\tH_\t, \hs{5}
\ell = (1-H) \tH_\t^{-1} \sin\t\cos\t, \nn
B^{(1)} &=& \frac{1-H}{H} dt \wedge dy_1 \cos\t, \hs{5}
B^{(2)} = \frac{1-H}{H} dt \wedge dy_1 \sin\t.
\ena
This agrees with ref.~\cite{RT}. All possible solutions thus seem to be
obtained without using boost for single charge case.

We can continue the procedure to generate more solutions, but we have already
generated enough examples to understand the general structure of solutions
of this kind.

\sect{Non-marginal solutions with two charges}

In this section, we proceed to the analysis of solutions obtained from
marginal solutions with two charges. We can introduce one angle at each
step of reduction and T-duality, thus producing general non-marginal bound
states of various marginal solutions. However, this produces quite complicated
solutions without much physical insight, and so below we present examples
of two-body bound states and the procedure how to obtain these. Some of
them are known ones, but others are new.

\subsection{Solutions obtained from $2_M\perp 2_M$}

Suppose that we start from $2_M\perp 2_M$ solution with the world-volume
coordinates
\bea
\begin{array}{ccl}
2_M & \perp & 2_M \vs{-3} \\
{\scriptstyle (y_1,y_2)} & & {\scriptstyle (y_3,y_4)} \hs{2},
\end{array}
\ena
where we have indicated the world-volume coordinates below each $2_M$-brane.
Making reduction in the directions $x_1$ or $y_4$, we get
\bea
\begin{array}{ccl}
(2_D & \perp & 2_D)_A \hs{2}, \vs{-3} \\
{\scriptstyle (y_1,y_2)} & & {\scriptstyle (y_3,y_4)}
\end{array} \hs{3}
\begin{array}{ccl}
(2_D & \perp & 1_F)_A \hs{2}, \vs{-3} \\
{\scriptstyle (y_1,y_2)} & & {\scriptstyle (y_4)}
\end{array}
\ena
or their bound state.

Applying the T-duality rule in all possible directions to the first solution,
we get
\bea
\begin{array}{ccl}
(3_D & \perp & 3_D)_B \hs{2}, \vs{-3} \\
{\scriptstyle (y_1,y_2,x_1)} & & {\scriptstyle (y_3,y_4,x_1)}
\end{array} \hs{3}
\begin{array}{ccl}
(3_D & \perp & 1_D)_B \hs{2}, \vs{-3} \\
{\scriptstyle (y_1,y_2,y_3)} & & {\scriptstyle (y_3)}
\end{array}
\label{sol1}
\ena
Similarly from the second solution, we get
\bea
\begin{array}{ccl}
(3_D & \perp & 0_w)_B \hs{2}, \vs{-3} \\
{\scriptstyle (y_1,y_2,y_3)} & & {\scriptstyle (y_3)}
\end{array} \hs{3}
\begin{array}{ccl}
(1_D & \perp & 1_F)_B \hs{2}, \vs{-3} \\
{\scriptstyle (y_1)} & & {\scriptstyle (y_3)}
\end{array} \hs{3}
\begin{array}{ccl}
(3_D & \perp & 1_F)_B \hs{2}, \vs{-3} \\
{\scriptstyle (y_1,y_2,x_1)} & & {\scriptstyle (y_3)}
\end{array}
\label{sol2}
\ena

We now apply T-duality to the rotated directions in all possible way to
the first solution in (\ref{sol1}) to find
\bea
T_{x_2-x_3} &:& (2_D \perp 2_D + 4_D \perp 4_D)_A, \nn
T_{x_2-y_2} &:& (2_D \perp 2_D + 2_D \perp 4_D)_A, \nn
T_{y_2-y_3} &:& (2_D \perp 4_D + 4_D \perp 2_D)_A, \nn
T_{y_2-x_3} &:& (2_D \perp 4_D + 4_D \perp 4_D)_A,
\ena
which can be lifted by the rule in the appendix to 11-dimensional non-marginal
solutions\footnote{The relative order of the fundamental solutions should be
noted. For example, the third solution is different from $(2_M\perp 5_M + 
2_M\perp 5_M)$ solution which is trivial in 11-dimensional sense.}
\bea
(2_M \perp 2_M + 5_M \perp 5_M), \nn
(2_M \perp 2_M + 2_M \perp 5_M), \nn
(2_M \perp 5_M + 5_M \perp 2_M), \nn
(2_M \perp 5_M + 5_M \perp 5_M).
\ena

{}From the second solution in (\ref{sol1}), similar procedure yields
the solutions\footnote{Here and in what follows, we display only those which
yield solutions other than those already listed.}
\bea
T_{y_1-y_4} &:& (2_D \perp 2_D + 4_D \perp 0_w)_A, \nn
T_{y_4-x_2} &:& (4_D \perp 0_D + 4_D \perp 2_D)_A,
\ena
which can be lifted to 11-dimensional non-marginal solutions
\bea
(2_M \perp 2_M + 5_M \perp 0_w), \hs{3}
(2_M \perp 5_M + 0_w \perp 5_M).
\ena

{}From the first solution in (\ref{sol2}), we obtain
\bea
T_{y_3-y_2} &:& (2_D \perp 1_F + 2_D \perp 0_w)_A, \nn
T_{y_1-x_1} &:& (2_D \perp 0_w + 4_D \perp 0_w)_A,
\ena
which can be lifted to 11-dimensional non-marginal solutions
\bea
(2_M \perp 2_M + 2_M \perp 0_w), \hs{3}
(2_M \perp 0_w + 5_M \perp 0_w).
\ena
Here we again have not shown other possible bound states which yields
already listed 11-dimensional solutions even though they are different
solutions in 10 dimensions, because they can be simply obtained from
the 11-dimensional solutions.

We have also examined other cases and the only new solutions that can be
obtained from the above is that from the second of (\ref{sol2}) as
\bea
T_{y_1-y_3} : (0_D \perp 1_F + 2_D \perp 0_w)_A,
\ena
which can be lifted to 11-dimensional non-marginal solutions
\bea
(0_w \perp 2_M + 2_M \perp 0_w).
\ena
As a bonus, we find the orthogonal intersection rules for wave and other
solutions: Wave can intersects with $2_M$ and $5_M$ over a string
(or can propagate on world-volume direction). This actually produces known
boosted solutions~\cite{TS1,GAR}.

As a check, we now summarize the explicit solutions listed above.
Of course, there are several routes to obtain same solutions and we show
below one possible way for each solutions.
Let us start from a rotated $2_M \perp 2_M$ solution
\bea
ds_{11}^2 &=& (H_1 H_2)^{1/3} \left[ - (H_1 H_2)^{-1} dt^2
 + H_1^{-1} ( dy_1^2 + dy_2^2 )  \right. \nn
&& \left. + H_2^{-1} \left( dy_3^2 + (dy_4 \cos\t
 + dx_1 \sin\t )^2 \right) + (- dy_4 \sin\t + dx_1 \cos\t )^2
 + \sum_{i=2}^6 dx_i^2 \right], \nn
C &=& \frac{1-H_1}{H_1} dt \wedge dy_1 \wedge dy_2 +
 \frac{1-H_2}{H_2} dt \wedge dy_3 \wedge (dy_4 \cos\t + dx_1 \sin\t ),
\label{22M}
\ena
where $H_1,H_2$ are harmonic functions depending on the transverse
coordinates $x_2, \cdots, x_6$.

Upon reduction in $y_4$ direction, we get type IIA $2_D \perp 1_F +
2_D \perp 2_D$ solution:
\bea
ds_{A}^2 &=& (H_1 \tH_2)^{1/2} \left[ - (H_1 H_2)^{-1} dt^2
 + H_1^{-1} ( dy_1^2 + dy_2^2 ) + H_2^{-1} dy_3^2 + \tH_2^{-1} dx_1^2
 + \sum_{i=2}^6 dx_i^2 \right], \nn
C &=& \frac{1-H_1}{H_1} dt \wedge dy_1 \wedge dy_2 +
 \frac{1-H_2}{H_2} dt \wedge dy_3 \wedge dx_1 \sin\t ,\hs{3}
e^{2\phi} = H_1^{1/2}H_2^{-1} \tH_2^{3/2}, \nn
B^{(1)} &=& \frac{1-H_2}{H_2} dt \wedge dy_3 \cos\t ,\hs{3}
A^{(1)} = \tH_2^{-1}(1-H_2)dx_1 \sin\t \cos\t.
\label{22}
\ena

\subsubsection{$2_M\perp 2_M$ and $5_M \perp 5_M$ bound state}

We start from (\ref{22}) with $\t=\frac{\pi}{2}$ and make the following
sequence of duality and lifting:
\bea
&& 2_M \perp 2_M \stackrel{R(\t=\frac{\pi}{2})}{\to} (2_D \perp 2_D)_A
 \stackrel{T_{x_2}}{\to} (3_D \perp 3_D)_B \stackrel{T_{x_2-x_3}}{\to}
 (2_D \perp 2_D + 4_D \perp 4_D)_A \nn
&& \hs{10} \stackrel{{\rm lift}(y_4)}{\to} (2_M\perp 2_M + 5_M \perp 5_M).
\label{2255}
\ena
The solution thus obtained is
\bea
ds_{11}^2 &=& (H_1 H_2 \tH_{12})^{1/3} \left[ - (H_1 H_2)^{-1} dt^2
 + \tH_{12}^{-1} ( dx_2^2 + dx_3^2 + dy_4^2 ) + H_1^{-1} (dy_1^2 + dy_2^2)
 \right. \nn
&& \left. \hs{10} + H_2^{-1} (dy_3^2 + dx_1^2)
 + \sum_{i=4}^6 dx_i^2 \right],\nn
dC &=& \left( dH_1^{-1} \wedge dt \wedge dy_1 \wedge dy_2
 + dH_2^{-1} \wedge dt \wedge dy_3 \wedge dx_1 \right) \cos\t \nn
&& \hs{10} + (*d H_1 \wedge dy_3 \wedge dx_1 + *d H_2 \wedge dy_1 \wedge dy_2)
 \sin\t \nn
&& \hs{10} + d\left(\frac{1-H_1 H_2}{\tH_{12}} \right)\wedge dx_2 \wedge dx_3
 \wedge dy_4 \cos\t \sin\t,
\label{s2255}
\ena
where the angle $\t$ is reintroduced in the angled duality in (\ref{2255}) and
\bea
\tH_{12} = \cos^2 \t + H_1 H_2 \sin^2 \t,
\label{h12}
\ena
in agreement with ref.~\cite{CC}. In that reference, the authors used
8-dimensional large symmetry $SL(2,R)$ to find this solution. However, we
do not have to refer to such a special symmetry realized only in lower
dimensions, but can derive these solutions step by step. Also this
step-by-step method can produce more general bound states.

\subsubsection{$2_M\perp (2_M + 5_M)$ bound state}

We make different T-duality to the solution $(3_D\perp 3_D)_B$ in
(\ref{2255}):
\bea
(3_D \perp 3_D)_B \stackrel{T_{x_2-y_2}}{\to}
 (2_D \perp 2_D + 2_D \perp 4_D)_A
 \stackrel{{\rm lift}(y_4)}{\to} (2_M\perp 2_M + 2_M \perp 5_M).
\label{2225}
\ena
The result is\footnote{Here and in the following, $*$ always means a dual
with respect to the transverse space. In the present case of eq.~(\ref{s2225}),
the transverse space consists of $x_i,i=3,\cdots,6$.}
\bea
ds_{11}^2 &=& (H_1 H_2 \tH_{2})^{1/3} \left[ - (H_1 H_2)^{-1} dt^2
 + H_1^{-1} dy_1^2 + (H_1 \tH_{2})^{-1} dy_2^2 + H_2^{-1} (dy_3^2 + dx_1^2)
 \right. \nn
&& \left. \hs{10} + \tH_2^{-1} (dy_4^2 + dx_2^2)
 + \sum_{i=3}^6 dx_i^2 \right],\nn
dC &=& d\left(\frac{1-H_2}{\tH_2}\right) \wedge dx_2 \wedge dy_2
 \wedge dy_4 \sin\t \cos\t + dH_1^{-1} \wedge dt \wedge dy_1 \wedge dy_2 \nn
&& \hs{5} + dH_2^{-1} \wedge dt \wedge dy_3 \wedge dx_1 \cos\t
 + *dH_2 \wedge dy_1 \sin\t,
\label{s2225}
\ena
This solution may be interpreted as an orthogonal intersection of
$2_M$ brane and the non-marginal solution $(2_M+5_M)$ in (\ref{s25}).
Viewed this way, this solution is discussed in ref.~\cite{C}. There the
author derived this solution by empirical rule that the solution should
agree with known orthogonal intersecting ones for $\t=0,\frac{\pi}{2}$.
This can be automatically generated by T-dualities.

\subsubsection{$2_M\perp 5_M$ and $5_M \perp 2_M$ bound state}

Another transformation to $(3_D \perp 3_D)_B$
\bea
(3_D \perp 3_D)_B \stackrel{T_{y_2-y_3}}{\to}
 (2_D \perp 4_D + 4_D \perp 2_D)_A
 \stackrel{{\rm lift}(y_4)}{\to} (2_M\perp 5_M + 5_M \perp 2_M),
\label{2552}
\ena
yields the solution
\bea
ds_{11}^2 &=& (H_1 H_2 {\hat H}_{12})^{1/3} \left[ (H_1 H_2)^{-1} (-dt^2
 + dx_2^2) + H_1^{-1} dy_1^2 + {\hat H}_{12}^{-1} (dy_2^2 + dy_3^2 +dy_4^2)
 \right. \nn
&& \left. \hs{5} + H_2^{-1} dx_1^2 + \sum_{i=3}^6 dx_i^2 \right],\nn
dC &=& d\left(\frac{H_2-H_1}{{\hat H}_{12}}\right) \wedge dy_2 \wedge dy_3
 \wedge dy_4 \sin\t \cos\t \nn
&& - \left[dH_1^{-1}\wedge dt \wedge dy_1\wedge dx_2
 + *dH_2 \wedge dy_1 \right] \cos\t \nn
&& - \left[dH_2^{-1}\wedge dt \wedge dx_1\wedge dx_2
 + *dH_2 \wedge dx_1 \right] \sin\t,
\label{s2552}
\ena
where
\bea
\hat H_{12} = H_1 \sin^2 \t + H_2 \cos^2 \t.
\label{hh12}
\ena
This solution is also derived by using 8-dimensional symmetry~\cite{CC}.

\subsubsection{$(2_M + 5_M) \perp 5_M$ bound state}

Another route is
\bea
(3_D \perp 3_D)_B \stackrel{T_{y_2-x_3}}{\to}
 (2_D \perp 4_D + 4_D \perp 4_D)_A
 \stackrel{{\rm lift}(y_4)}{\to} (2_M\perp 5_M + 5_M \perp 5_M).
\label{2555}
\ena
The resulting solution is
\bea
ds_{11}^2 &=& (H_1 \tH_{1})^{1/3} H_2^{2/3} \left[ (H_1 H_2)^{-1}(- dt^2
 + dx_2^2) + H_1^{-1} dy_1^2 + (H_2 \tH_{1})^{-1} (dy_2^2 + dy_4^2)
 \right. \nn
&& \left. \hs{10}  + \tH_1^{-1} dx_3^2 + H_2^{-1} (dy_3^2 + dx_1^2)
 + \sum_{i=4}^6 dx_i^2 \right],\nn
dC &=& d\left(\frac{1-H_1}{\tH_1}\right) \wedge dy_2 \wedge dx_3
 \wedge dy_4 \sin\t \cos\t - dH_1^{-1} \wedge dt \wedge dy_1
 \wedge dx_2 \cos\t \nn
&& \hs{5} + *dH_1 \wedge dy_3 \wedge dx_1 \sin\t
 + *dH_2 \wedge dy_1 \wedge dx_3.
\label{s2555}
\ena
This again may be regarded as an orthogonal intersection of $(2_M+ 5_M)$
in (\ref{s25}) and $5_M$.

\subsubsection{$2_M\perp (0_w + 2_M)$ bound state}

{}From (\ref{22}) with $\t=0$, we find a new solution by
\bea
&& (2_D \perp 1_F)_A \stackrel{T_{y_1}}{\to} (3_D \perp 0_w)_B
 \stackrel{T_{y_2-y_3}}{\to} (2_D\perp 0_w + 2_D \perp 1_F)_A \nn
&& \hs{10} \stackrel{{\rm lift}(y_4)}{\to} (2_M\perp 0_w + 2_M \perp 2_M).
\label{2022}
\ena
The solution is
\bea
ds_{11}^2 &=& (H_1 \tH_{2})^{1/3} \left[ (H_1\tH_2)^{-1}\left(- dt^2 + dy_3^2
 + (H_2-1) (dt\cos\t + dy_3)^2 \right) + H_1^{-1} dy_1^2 \right. \nn
&& \left. \hs{10}  + \tH_2^{-1} (dy_2^2 + dy_4^2)
 + \sum_{i=1}^6 dx_i^2 \right],\nn
C &=& \frac{1-H_1}{H_1} dt \wedge dy_3 \wedge dy_1 +
 \frac{1-H_2}{\tH_2} dy_2 \wedge ( dt + dy_3 \cos\t) \wedge dy_4 \sin\t.
\label{s2022}
\ena
This may be regarded as an orthogonal intersection of $2_M$ and
$(0_w+2_M)$ in (\ref{s02}).

\subsubsection{$2_M\perp 2_M$ and $5_M \perp 0_w$ bound state}

{}From $(3_D \perp 0_w)_B$ in (\ref{2022}), we obtain
\bea
(3_D \perp 0_w)_B \stackrel{T_{y_3-x_1}}{\to}
 (2_D\perp 1_F + 4_D \perp 0_w)_A
 \stackrel{{\rm lift}(y_4)}{\to} (2_M\perp 2_M + 5_M \perp 0_w).
\label{2250}
\ena
The solution is
\bea
ds_{11}^2 &=& (H_1 {\hat H}_{12})^{1/3} \left[
 -(H_1{\hat H}_{12})^{-1} \tH_1 dt^2 + H_1^{-1} (dy_1^2 + dy_2^2) 
 \right. \nn
&& \left. \hs{5} + {\hat H}_{12}^{-1}\left( (H_2-1) (dt \sin\t + dx_1)^2
 + dy_3^2 + dy_4^2 + dx_1^2 \right) + \sum_{i=2}^6 dx_i^2 \right],\nn
dC &=& d\left(\frac{H_2-H_1}{\hat H_{12}}\right)\wedge dy_3 \wedge dx_1
 \wedge dy_4 \sin\t \cos\t + d\left(\frac{1-H_2}{\hat H_{12}}\right)
 \wedge dt \wedge dy_3 \wedge dy_4 \cos\t \nn
&& \hs{5} + dH_1^{-1}\wedge dt \wedge dy_1 \wedge dy_2 \cos\t
 + *d H_1 \sin\t,
\label{s2250}
\ena
where ${\hat H_{12}}$ is defined in (\ref{hh12}). This solution has
a special feature that it contains both $\hat H_{12}$ and $\tH_1$ in the
metric, and is not a simple one obtained by introducing a wave in the null
isometry direction as in ref.~\cite{GAR}.

\subsubsection{$(2_M + 5_M) \perp 0_w$ bound state}

Another new solution is derived by
\bea
(3_D \perp 0_w)_B \stackrel{T_{y_1-x_1}}{\to}
 (2_D\perp 0_w + 4_D \perp 0_w)_A
 \stackrel{{\rm lift}(y_4)}{\to} (2_M\perp 0_w + 5_M \perp 0_w).
\label{2050}
\ena
The solution is
\bea
ds_{11}^2 &=& (H_1 {\tH}_{1})^{1/3} \left[ H_1^{-1}\left(- dt^2 + dy_2^2
 + dy_3^2 + (H_2-1) (dt + dy_3)^2 \right) \right. \nn
&& \left. \hs{10}  + {\tH}_{1}^{-1} ( dy_1^2 + dy_4^2 + dx_1^2)
 + \sum_{i=2}^6 dx_i^2 \right],\nn
dC &=& - dH_1^{-1} \wedge dt \wedge dy_3 \wedge dy_2 \cos\t
 + *dH_1 \sin\t \nn
&& \hs{10} - d\tH_{1}^{-1}\wedge dx_1 \wedge dy_1 \wedge dy_4 \cot\t.
\label{s2050}
\ena
This may be considered an orthogonal intersection of $(2_M+5_M)$ in
(\ref{s25}) and $0_w$. In fact, this can also be obtained from (\ref{s25})
just by introducing wave in the null isometry direction \cite{GAR}.

\subsubsection{$0_w\perp 2_M$ and $2_M \perp 0_w$ bound state}

Again from (\ref{22}) with $\t=0$, we get
\bea
&& (2_D \perp 1_F)_A \stackrel{T_{y_2}}{\to} (1_D\perp 1_F)_B
 \stackrel{T_{y_1-y_3}}{\to} (0_D\perp 1_F + 2_D \perp 0_w)_A \nn
&& \hs{10} \stackrel{{\rm lift}(y_4)}{\to} (0_w\perp 2_M + 2_M \perp 0_w).
\label{0220}
\ena
The solution is
\bea
ds_{11}^2 &=& {\hat H}_{12}^{1/3} \left[ {\hat H}_{12}^{-1}\left(
 - dt^2 + (H_2-1) (dt \sin\t-dy_1)^2 + (H_1-1)(dt\cos\t-dy_4)^2
 \right. \right. \nn
&& \left. \left. \hs{5} + dy_1^2 + dy_3^2 + dy_4^2 \right)
 + dy_2^2 + \sum_{i=1}^6 dx_i^2 \right],\nn
C &=& \frac{H_1-1}{\hat H_{12}} (dt-dy_4\cos\t) \wedge dy_1 \wedge dy_3 \sin\t
\nn \hs{3} &&
 + \frac{1-H_2}{\hat H_{12}} (dt-dy_1\sin\t) \wedge dy_3 \wedge dy_4 \cos\t.
\label{s0220}
\ena

\subsubsection{$5_M\perp 0_w$ and $5_M \perp 2_M$ bound state}

{}From $(2_D\perp 1_F)$, we also get
\bea
&& (2_D \perp 1_F)_A \stackrel{T_{x_1}}{\to} (3_D\perp 1_F)_B
 \stackrel{T_{y_3-x_2}}{\to} (4_D\perp 0_w + 4_D \perp 1_F)_A \nn
&& \hs{10} \stackrel{{\rm lift}(y_4)}{\to} (5_M\perp 0_w + 5_M \perp 2_M).
\label{5052}
\ena
The solution is
\bea
ds_{11}^2 &=& H_1^{2/3} \tH_2^{1/3} \left[ (H_1\tH_2)^{-1}\left(- dt^2
 + (H_2-1) (dt\cos\t + dy_3)^2 + dy_3^2 + dy_4^2 \right) \right. \nn
&& \hs{5} \left. + H_1^{-1} (dy_1^2 + dy_2^2 + dx_1^2)
 + \tH_2^{-1} dx_2^2 + \sum_{i=3}^6 dx_i^2 \right],\nn
dC &=& d(\tH_2 \sin\t)^{-1} \wedge (dt+dy_3 \cos\t) \wedge dx_2 \wedge dy_4
 - * dH_1 \wedge dx_2.
\label{s5052}
\ena
This is again an orthogonal intersection of $5_M$ and $(0_w+2_M)$ in
(\ref{s02}).

This completes the two-body new solutions obtained from $2_M\perp 2_M$
just by a single step of going through type IIB solutions. We now turn to
bound states obtained from other marginal solutions.

\subsection{Solutions obtained from $5_M\perp 5_M$}

If we start from $5_M\perp 5_M$ solution with the world-volume
coordinates
\bea
\begin{array}{ccc}
5_M & \perp & 5_M \vs{-3} \\
{\scriptstyle (y_1,\cdots,y_5)} & & {\scriptstyle (y_1,\cdots,y_3,y_6,y_7)}
\hs{2},
\end{array}
\ena
we get upon reduction in the directions $y_3,y_4$ or $x_1$
\bea
(4_D \perp 4_D)_A, \hs{2}
(4_D \perp 5_S)_A, \hs{2}
(5_S \perp 5_S)_A,
\ena
or their bound state.

Applying the T-duality rule in all possible directions, we get
\bea
(3_D \perp 3_D)_B, \hs{2}
(3_D \perp 5_D)_B, \hs{2}
(5_D \perp 5_D)_B, \hs{2}
(3_D \perp 5_S)_B, \hs{2}
(3_D \perp 0_m)_B, \nn
(5_D \perp 5_S)_B, \hs{2}
(5_D \perp 0_m)_B, \hs{2}
(5_S \perp 5_S)_B, \hs{2}
(5_S \perp 0_m)_B, \hs{2}
(0_m \perp 0_m)_B,
\ena
and their bound states. From these, after T-duality we find new solutions
\bea
(2_D \perp 4_D + 2_D \perp 6_D)_A, \hs{5}
(2_D \perp 4_D + 4_D \perp 6_D)_A, \nn
(2_D \perp 6_D + 4_D \perp 4_D)_A, \hs{5}
(2_D \perp 6_D + 4_D \perp 6_D)_A, \nn
(4_D \perp 4_D + 4_D \perp 6_D)_A, \hs{5}
(4_D \perp 4_D + 6_D \perp 6_D)_A, \nn
(4_D \perp 6_D + 6_D \perp 4_D)_A, \hs{5}
(4_D \perp 6_D + 6_D \perp 6_D)_A,
\ena
which yield the following new solutions:
\bea
(2_M \perp 5_M + 2_M \perp 0_m), \hs{5}
(2_M \perp 5_M + 5_M \perp 0_m), \nn
(2_M \perp 0_m + 5_M \perp 5_M), \hs{5}
(2_M \perp 0_m + 5_M \perp 0_m), \nn
(5_M \perp 5_M + 5_M \perp 0_m). \hs{5}
(5_M \perp 5_M + 0_m \perp 0_m), \nn
(5_M \perp 0_m + 0_m \perp 5_M), \hs{5}
(5_M \perp 0_m + 0_m \perp 0_m).
\ena
As a by-product of this procedure, we can read off the orthogonal
intersection rules for monopole and other solutions and itself, which are
discussed in a recent paper~\cite{BR}. We note in particular that there
are two possible intersections of two monopoles arising from that through
$(6_D\perp 6_D)_A$ and that obtained from $(5_S\perp 5_S)_A$.\footnote{
Both are over 4-branes, but how the gauge fields are introduced is different.}
{}For the orthogonal intersection of monopole and $2_M$ and $5_M$ branes,
there are also other intersections that can be derived through different
route from $2_M + 5_M$~\cite{BR}. The explicit
form of the solutions will be given below.

Solutions discussed in this subsection are all new bound states.

\subsubsection{Solution $(5_M + 0_m) \perp 2_M$}

We make the following transformation:
\bea
&&(5_M\perp 5_M) \stackrel{R_{y_3}}{\to} (4_D\perp 4_D)_A
 \stackrel{T_{y_6}}{\to} (5_D\perp 3_D)_B \stackrel{T_{y_2-y_7}}{\to}
 (4_D\perp 2_D + 6_D \perp 2_D)_A \nn
&& \hs{10} \stackrel{{\rm lift}(z)}{\to} (5_M \perp 2_M + 0_m \perp 2_M).
\label{2520}
\ena
The solution is
\bea
ds_{11}^2 &=& H_1^{2/3}(\tH_1 H_2)^{1/3} \left[(H_1H_2)^{-1}(
 - dt^2 + dy_1^2) + \tH_1^{-1} dy_2^2 + H_1^{-1}(
 dy_4^2 + dy_5^2 + dy_6^2) \right. \nn
&& \hs{5} \left. + (\tH_1 H_2)^{-1} dy_7^2 + (H_1\tH_1)^{-1}(dz
 + A_i \sin\t dx_i)^2 + \sum_{i=1}^3 dx_i^2 \right], \nn
dC &=& dH_2^{-1} \wedge dt \wedge dy_1 \wedge dy_7 + d(\tH_1^{-1} A_i dx_i)
 \wedge dy_2 \wedge dy_7 \cos\t \nn
&& \hs{5} + d\tH_1^{-1} \wedge dy_2 \wedge dy_7 \wedge dz \cot\t,\nn
dA &=& *dH_1,
\label{s2520}
\ena
where $\tH_1$ is as defined in (\ref{ht}) with $H$ replaced by $H_1$.\footnote{
In the last line of (\ref{s2520}), $A$ is a 1-form $A_i dx_i$. The
same notation is used in the following as well.}
This solution may be understood as the orthogonal intersection of the bound
state $(5_M+0_m)$ given in (\ref{sm5}) and  $2_M$.

\subsubsection{Solution $5_M \perp 2_M + 0_m \perp 5_M$}

{}From $(5_D\perp 3_D)_B$ in (\ref{2520}), we proceed as
\bea
(5_D\perp 3_D)_B \stackrel{T_{y_2-x_1}}{\to}
 (4_D\perp 2_D + 6_D \perp 4_D)_A
 \stackrel{{\rm lift}(z)}{\to} (5_M \perp 2_M + 0_m \perp 5_M).
\ena
to find
\bea
ds_{11}^2 &=& H_1^{2/3}(\tH_{12} H_2)^{1/3} \left[(H_1H_2)^{-1}(
 - dt^2 + dy_1^2) + \tH_{12}^{-1} (dy_2^2 + dx_1^2)
 + H_1^{-1}( dy_4^2 + dy_5^2 \right. \nn
&& \hs{5} \left. + dy_6^2)
 + H_2^{-1} dy_7^2 + (H_1\tH_{12})^{-1}(dz + A_{y_7} \sin\t dy_7)^2
 + \sum_{i=2}^3 dx_i^2 \right], \nn
dC &=& \left[d(\tH_{12}^{-1}A_{y_7}) \wedge dy_2 \wedge dy_7 \wedge dx_1
 + dH_2^{-1} \wedge dt \wedge dy_1 \wedge dy_7 \right]\cos\t \nn
&& \hs{5} + *dH_2 \wedge dy_4 \wedge dy_5 \wedge dy_6 \sin\t
 + d\tH_{12}^{-1} \wedge dy_2 \wedge dx_1 \wedge dz \cot\t, \nn
dA &=& - *dH_1 \wedge dy_7,
\label{s255m}
\ena
where $\tH_{12}$ is defined in (\ref{h12}).

\subsubsection{Solution $5_M \perp 5_M + 0_m \perp 2_M$}

Another transformation on $(5_D\perp 3_D)_B$ yields
\bea
(5_D\perp 3_D)_B \stackrel{T_{y_6-y_7}}{\to}
 (4_D \perp 4_D + 6_D\perp 2_D)_A
 \stackrel{{\rm lift}(z)}{\to} (5_M \perp 5_M + 0_m \perp 2_M).
\ena
The solution is
\bea
ds_{11}^2 &=& H_1^{2/3}({\hat H}_{12} H_2)^{1/3} \left[(H_1H_2)^{-1}(
 - dt^2 + dy_1^2 + dy_2^2) + H_1^{-1} (dy_4^2 + dy_5^2) \right. \nn
&& \hs{5} \left. + {\hat H}_{12}^{-1}( dy_6^2 + dy_7^2)
 + (H_1{\hat H}_{12})^{-1}(dz+A_i \sin\t dx_i)^2
 + \sum_{i=1}^3 dx_i^2 \right], \nn
dC &=& \left[d(A_i dx_i H_2 {\hat H}_{12}^{-1}) \wedge dy_6 \wedge dy_7
 + *dH_2 \wedge dy_4 \wedge dy_5 \right] \cos\t
 \nn
&& \hs{5} + dH_2^{-1} \wedge dt \wedge dy_1 \wedge dy_2 \sin\t
 + d\left(\frac{H_2-H_1}{{\hat H}_{12}}\right) \wedge dy_6
 \wedge dy_7 \wedge dz \sin\t \cos\t, \nn
dA &=& *dH_1,
\label{s2m55}
\ena
where $\hat H_{12}$ is given in (\ref{hh12}).

\subsubsection{Solution $0_m \perp (2_M + 5_M)$}

Another solution is obtained by
\bea
(5_D\perp 3_D) \stackrel{T_{y_7-x_1}}{\to}
 (6_D\perp 2_D + 6_D \perp 4_D)_A
 \stackrel{{\rm lift}(z)}{\to} (0_m \perp 2_M + 0_m \perp 5_M).
\ena
The result is
\bea
ds_{11}^2 &=& (\tH_2 H_2)^{1/3} \left[H_2^{-1}(
 - dt^2 + dy_1^2 + dy_2^2) + dy_4^2 + dy_5^2 + dy_6^2 \right. \nn
&& \hs{5} \left. + \tH_2^{-1} dy_7^2 + H_1 \tH_2^{-1} dx_1^2
 + (H_1 \tH_2)^{-1}(dz+A_i dx_i)^2 + H_1 \sum_{i=2}^3 dx_i^2 \right], \nn
dC &=& *dH_2 \wedge dy_4 \wedge dy_5 \wedge dy_6 \sin\t
 + dH_2^{-1} \wedge dt \wedge dy_1 \wedge dy_2 \cos\t \nn
&& \hs{5} + d\tH_2^{-1} \wedge dy_7\wedge dx_1 \wedge dz \cot\t, \nn
dA &=& - *dH_1 \wedge dx_1.
\label{s2m5m}
\ena
This may be understood as an orthogonal intersection of $0_m$ and
$(2_M + 5_M)$ in (\ref{s25}).

{}From the solutions discussed so far, we can read off the orthogonal
intersection rules for the monopole with $5_M$ and $2_M$; the intersections
are over 2- and 5-branes, respectively, in the above solutions. (There are
also other cases obtained from $2_M \perp 5_M$.)

\subsubsection{Solution $(5_M + 0_m) \perp 5_M$}

The last one from $(5_D\perp 3_D)_B$ is
\bea
(5_D\perp 3_D)_B \stackrel{T_{y_6-x_1}}{\to}
 (4_D\perp 4_D + 6_D \perp 4_D)_A
 \stackrel{{\rm lift}(z)}{\to} (5_M \perp 5_M + 0_m \perp 5_M).
\ena
The solution is
\bea
ds_{11}^2 &=& \hs{-3}(H_1 H_2)^{2/3} \tH_1^{1/3} \left[(H_1 H_2)^{-1}(
 - dt^2 + dy_1^2 + dy_2^2) + H_1^{-1} ( dy_4^2 + dy_5^2)
 + (\tH_1 H_2)^{-1} dy_6^2 \right. \nn
&& \hs{2} \left. + H_2^{-1} dy_7^2 + \tH_1^{-1} dx_1^2
 + (H_1 H_2 \tH_1)^{-1}(dz+A_{y_7} \sin\t dy_7)^2
 + \sum_{i=2}^3 dx_i^2 \right], \nn
dC &=& d(A_i dx_i \tH_1^{-1}) \wedge dy_6 \wedge dy_7 \wedge dx_1 \cos\t
 + *dH_2 \wedge dy_4 \wedge dy_5 \wedge dx_1 \nn
&& \hs{5} + d\tH_1^{-1} \wedge dy_6 \wedge dx_1 \wedge dz \cot\t, \nn
dA &=& - *dH_1 \wedge dy_7.
\label{s555m}
\ena
This may be regarded as an orthogonal intersection of $(5_M + 0_m)$ in
(\ref{sm5}) and  $5_M$.

\subsubsection{Solution $5_M \perp 5_M + 0_m \perp 0_m$}

{}From $(4_D\perp 4_D)_A$ in (\ref{2520}), we take the route
\bea
&& (4_D\perp 4_D)_A \stackrel{T_{x_1}}{\to} (5_D\perp 5_D)_B
 \stackrel{T_{x_1-x_2}}{\to} (4_D\perp 4_D + 6_D \perp 6_D)_A \nn
&& \hs{3} \stackrel{{\rm lift}(z)}{\to} (5_M \perp 5_M + 0_m \perp 0_m).
\label{5500}
\ena
The solution is
\bea
ds_{11}^2 &=& (H_1 H_2)^{2/3} \tH_{12}^{1/3} \left[(H_1H_2)^{-1}(
 - dt^2 + dy_1^2 + dy_2^2 ) + \tH_{12}^{-1}( dx_1^2 + dx_2^2)
 + H_1^{-1} ( dy_4^2 \right. \nn
&& \left. \hs{-5} + dy_5^2) + H_2^{-1} ( dy_6^2 + dy_7^2)
 + (H_1 H_2 \tH_{12})^{-1}\left\{dz+\sin\t (A_{y_5} dy_5 + B_{y_7} dy_7)
 \right\}^2 + dx_3^2 \right], \nn
dC &=& d\left[ (A_{y_5} dy_5 + B_{y_7} dy_7) \tH_{12}^{-1}\right]
 \wedge dx_1 \wedge dx_2 \cos\t \nn
&& \hs{5} + d\left(\frac{1-H_1 H_2}{\tH_{12}}\right) \wedge dx_1
 \wedge dx_2 \wedge dz \sin\t \cos\t, \nn
dA_{y_5} &=& \pa_{x_3} H_2 dy_4, \hs{5}
dB_{y_7} = \pa_{x_3} H_1 dy_6,
\ena
where $\tH_{12}$ is defined in (\ref{h12}) and the monopole gauge field
1-forms $A$ and $B$ live in the spaces $(x_3,y_4,y_5)$ and $(x_3,y_6,y_7)$,
respectively. Here we have chosen the gauge $A_{x_3} = A_{y_6}=B_{x_3}=
B_{y_4}=0$ (this is possible because the harmonic functions only depend on
the coordinate $x_3$). For $\t=\frac{\pi}{2}$, this agrees with the orthogonal
intersection for monopoles given in~\cite{BR}:
\bea
ds_{11}^2 &=& - dt^2 + dy_1^2 + dy_2^2 + dx_1^2 + dx_2^2
 + H_2 ( dy_4^2 + dy_5^2) + H_1 ( dy_6^2 + dy_7^2) \nn
&& \hs{-5} + (H_1 H_2)^{-1}(dz+ A_{y_5} dy_5 + B_{y_7} dy_7)^2
 + H_1 H_2 dx_3^2.
\ena
For completeness, we also record another possible intersection obtained
from $5_S+5_S$:
\bea
ds_{11}^2 &=& - dt^2 + dy_1^2 + \cdots + dy_4^2 + H_1 dy_5^2
 + H_2 dy_6^2 + H_1^{-1} ( dz_1 + \sum_{i=1}^3 A_i dy_i)^2 \nn
&& \hs{-5} + H_2^{-1} ( dz_2 + \sum_{i=1}^3 B_i dy_i)^2
 + H_1 H_2 (dx_1^2 + dx_2^2),
\ena
where the harmonic functions depend on $x_1$ and $x_2$.

\subsubsection{Solution $5_M \perp 0_m + 0_m \perp 5_M$}

Another from $(5_D\perp 5_D)_B$ in (\ref{5500}) is
\bea
(5_D\perp 5_D)_B \stackrel{T_{y_5-y_7}}{\to} (4_D\perp 6_D + 6_D \perp 4_D)_A
 \stackrel{{\rm lift}(z)}{\to} (5_M \perp 0_m + 0_m \perp 5_M).
\ena
The solution is
\bea
ds_{11}^2 &=& (H_1 H_2)^{2/3} {\hat H}_{12}^{1/3} \left[(H_1 H_2)^{-1}(
 - dt^2 + dy_1^2 + dy_2^2 + dx_1^2) + H_1^{-1} dy_4^2
 \right. \nn
&& \hs{3} + H_2^{-1} dy_6^2 + {\hat H}_{12}^{-1} ( dy_5^2 + dy_7^2) \nn
&& \hs{3} \left. + (H_1 H_2 {\hat H}_{12})^{-1}(dz-\cos\t A_{y_4} dy_4
 + \sin\t B_{y_6} dy_6)^2 + \sum_{i=2}^3 dx_i^2 \right], \nn
dC &=& d\left[ \left\{\sin\t H_1 (A_{y_4} dy_4 - dz\cos\t)
 + \cos\t H_2 (B_{y_6} dy_6 + dz \sin\t) \right\} {\hat H}_{12}^{-1} \right]
 \nn&& \hs{5} \wedge dy_5 \wedge dy_7, \nn
dA_{y_4} &=& - *dH_2, \hs{5}
dB_{y_6} = - *dH_1,
\ena
where we have made a gauge choice similar to the above case.

\subsubsection{Solution $0_m \perp (5_M + 0_m)$}

The last solution in this series is obtained by
\bea
(5_D\perp 5_D)_B \stackrel{T_{y_7-x_1}}{\to} (6_D\perp 4_D + 6_D \perp 6_D)_A
 \stackrel{{\rm lift}(z)}{\to} (0_m \perp 5_M + 0_m \perp 0_m).
\ena
The solution is
\bea
ds_{11}^2 &=& H_2^{2/3} \tH_2^{1/3} \left[ H_2^{-1}(
 - dt^2 + dy_1^2 + dy_2^2 + dx_1^2) + dy_4^2 + dy_5^2
 + H_1 H_2^{-1} dy_6^2 \right. \nn
&& \hs{-3} \left. + \tH_2^{-1} dy_7^2 + H_1 \tH_2^{-1} dx_2^2
 + (H_1 H_2 \tH_2)^{-1}(dz+\sin\t A_{y_4} dy_4 + B_{y_6} dy_6)^2
 + H_1 dx_3^2 \right], \nn
dC &=& d\left[ (\sin\t A_{y_4}dy_4 + B_{y_6} dy_6 +dz) \tH_2^{-1}\right]
 \wedge dy_7 \wedge dx_2 \cot\t, \nn
dA_{y_4} &=& - \pa_{x_3}H_2 dy_5, \hs{5}
dB_{y_6} = - \pa_{x_3}H_1 dx_2.
\label{s5mmm}
\ena
This can be understood as an orthogonal intersection of $0_m$ and $(5_M+0_m)$
in (\ref{sm5}).

\subsection{Solutions obtained from $2_M\perp 5_M$}

We summarize solutions obtained from $2_M\perp 5_M$. Consider
\bea
\begin{array}{ccc}
2_M & \perp & 5_M \vs{-3} \\
{\scriptstyle (y_1,y_2)} & & {\scriptstyle (y_1,y_3,\cdots,y_6)}
\hs{2},
\end{array}
\ena
and we get upon reduction in the directions $y_1,y_2,y_3$ or $x_1$
\bea
(1_F \perp 4_D)_A, \hs{2}
(1_F \perp 5_S)_A, \hs{2}
(2_D \perp 4_D)_A, \hs{2}
(2_D \perp 5_S)_A,
\ena
or their bound state.

Applying the T-duality rule in all possible directions, we get
\bea
(0_w \perp 5_D)_B, \hs{2}
(1_F \perp 5_D)_B, \hs{2}
(0_w \perp 5_S)_B, \hs{2}
(1_F \perp 5_S)_B, \nn
(1_F \perp 0_m)_B, \hs{2}
(1_D \perp 5_D)_B, \hs{2}
(1_D \perp 5_S)_B, \hs{2}
(1_D \perp 0_m)_B,
\ena
and their bound states. From these, after T-duality we find new solutions
\bea
(1_F \perp 4_D + 0_w \perp 6_D)_A, \hs{5}
(0_w \perp 4_D + 0_w \perp 6_D)_A, \nn
(0_w \perp 6_D + 1_F \perp 6_D)_A, \hs{5}
(0_w \perp 5_S + 1_F \perp 0_m)_A,
\ena
which yield the following new solutions in 11 dimensions:
\bea
(2_M \perp 5_M + 0_w \perp 0_m), \hs{5}
(0_w \perp 5_M + 0_w \perp 0_m), \nn
(0_w \perp 0_m + 2_M \perp 0_m), \hs{5}
(0_w \perp 5_M + 2_M \perp 0_m).
\ena
We refrain from giving explicit metrics since it is straightforward
to derive them once the routes and possible solutions are given.

\subsection{Other bound state solutions}

It is clear that we can obtain many bound states by repeating
the above procedure starting from other possible solutions. The bound
states thus obtained include any two combinations of the solutions
\bea
&& 0_w \perp 2_M, \hs{2}
0_w \perp 5_M, \hs{2}
0_w \perp 0_m, \hs{2}
2_M \perp 2_M, \hs{2}
2_M \perp 5_M, \nn
&& 2_M \perp 0_m, \hs{2}
5_M \perp 5_M, \hs{2}
5_M \perp 0_m, \hs{2}
0_m \perp 0_m.
\ena

If we also include angles at each steps of reductions and dualities,
we can have more general non-marginal solutions interpolating these
solutions.

In the above examples, we have not continued making dualities to produce
more solutions. However, it is not our purpose here to exhaust these bound
states but to present examples of typical solutions and a systematic method
of producing general non-marginal solutions which we believe are useful
in searching for the construction rules for how to construct non-marginal
solutions directly in 11 dimensions. We will make an attempt to formulate
it in sect.~5.

Starting from three-charge solutions like $2_M\perp 2_M \perp 2_M$,
we can similarly construct bound states of these solutions involving
orthogonal intersections of all fundamental solutions summarized in
the introduction. Since the technique is now fairly clear, we do not
give explicit examples of these cases.

\sect{Solutions with tilted branes}

We have considered only reductions and T-dualities with rotations among
space coordinates. It is then natural to consider more general reductions
including boost which mixes time and space coordinates~\cite{RT,CP,BC,CC}.
Let us discuss the effect of boost in some detail since there is not much
discussion.\footnote{There are also solutions given in ref.~\cite{BM2},
but they reduce to orthogonal intersections for single center case~\cite{MM},
and are not discussed here.} We will show that the resulting solutions are
bound states with waves, some of which can also be obtained by the T-duality
transformations in previous sections. However, we have not considered further
T-duality transformations of these solutions. We now show that T-duality on
such solutions introduces titling of branes.

\subsection{Single brane}

We consider the $2_M$ brane~(\ref{2M}) and introduce the boost along
one of the transverse direction
\bea
t \to t \cosh \b - x_1 \sinh\b, \hs{5}
x_1 \to - t \sinh \b + x_1 \cosh\b.
\ena
(Boost on the world-volume direction is trivial for single brane.)
The metric $g_{x_1 x_1}$ becomes
\bea
g_{x_1 x_1} = H^{-2/3}(-\sinh^2 \b + H \cosh^2 \b).
\ena
This is well-defined for $\b=0$, but not for $\b=\infty$. To make both limits
well-defined, we should keep $Q \cosh^2\b$ finite. A convenient way to do this
is to replace the harmonic function $H$ in our solution by $\tH$ and also
put
\bea
\cosh\b=\frac{1}{\sin\t}, \hs{5}
\sinh\b=\frac{\cos\t}{\sin\t}.
\label{boost}
\ena
Namely we reduce the charge according to the boost. One then finds
\bea
ds_{11}^2 &=& {\tH_\t}^{1/3} \left[ \tH_\t^{-1} \left( -dt^2
 + dx_1^2 + (H-1)(dt \cos \t + dx_1)^2 +dy_1^2 + dy_2^2 \right) \right.\nn
&& \left. \hs{10}+ \sum_{i=2}^8 dx_i^2 \right], \nn
C &=& \frac{(1-H)\sin\t}{\tH_\t} (dt - dx_1 \cos \t)
\wedge dx_2 \wedge dx_1 ,
\label{ss02}
\ena
basically the same solution as (\ref{s02}), a bound state of wave and
$2_M$-brane. This is the general feature and one always gets bound states
with wave if one starts from other solutions. In fact, from $5_M$-brane,
one finds $(0_w+5_M)$ in (\ref{s05}), and from wave one has $(0_w+0_w)$:
\bea
ds_{11}^2 &=& -dt^2 + dy_1^2 + dy_2^2 + (H-1)(dt + \cos\t dy_1+\sin\t dy_2)^2
 + \sum_{i=1}^8 dx_i^2.
\ena
Since such bound states can also be obtained by angled T-duality and lifting
discussed in sect.~2, we do not need to consider such a boosted reduction.

We can consider further T-duality of these solutions. For example, we can make
reduction to (\ref{ss02}) in $x_1$ to get a type IIA solution,
which can be interpreted as a bound state of $(0_D+2_D)_A$~\cite{CP}.
{}For general angle $\t$ we have background $B^{(1)}$ which will produce
off-diagonal metrics after T-duality. However, we find that such off-diagonal
metrics can be removed by space rotations in 11 dimensions, and the solution
is equivalent to a trivial rotated $2_M+2_M$ bound state in (\ref{r2M}).
This is to be expected since the boost has the effect of rotating objects
lying in transverse direction with respect to it, but such a rotation of a
single brane can be removed by a coordinate rotation.

We can repeat the same analysis for $5_M$ brane~(\ref{5M}).
After the reduction, one gets a bound state of $(0_D+5_S)_A$~\cite{CP}.
One finds in this case again similar bound state solutions with apparent
off-diagonal metrics which can be removed.

\subsection{Tilted $(2_M + 2_M\perp 2_M)$ brane}

Starting from the $2_M\perp 2_M$ solution
\bea
ds_{11}^2 &=& (\tH_1 \tH_2)^{1/3} \left[ - (\tH_1 \tH_2)^{-1} dt^2
 + \tH_1^{-1} (dy_1^2 + dy_2^2) + \tH_2^{-1} (dy_3^2 + dy_4^2)
 + \sum_{i=1}^6 dx_i^2 \right], \nn
C &=& \frac{1-\tH_1}{\tH_1} dt \wedge dy_1 \wedge dy_2
 + \frac{1-\tH_2}{\tH_2} dt \wedge dy_3 \wedge dy_4,
\ena
there are two possibilities to introduce boost in 11 dimensions.
The first one is to do it in the transverse direction $x_1$ which
is discussed in ref.~\cite{CC}, and this again gives a bound state of
$(0_w+2_M \perp 2_M)$. Making reduction in $x_1$,
T-duality twice along $y_1$ and $y_3$, one gets a tilted $(2_D+ 2_D\perp 2_D)$
solution. We lift it to 11-dimensions to find tilted $(2_M+2_M\perp 2_M)$
solution
\bea
ds_{11}^2 &=& H_{12}^{1/3} \left[ - H_{12}^{-1} dt^2
 + (H_{12} \tH_1)^{-1} \left(\tH_1 dy_1 + (H_1-1) \sin\t \cos\t dy_2\right)^2
  + \tH_1^{-1} dy_2^2 \right. \nn
&& \left. + (H_{12} \tH_2)^{-1} \left(\tH_2 dy_3
 + (H_2-1) \sin\t \cos\t dy_4\right)^2
 + \tH_2^{-1} dy_4^2 + \sum_{i=1}^6 dx_i^2 \right], \nn
C &=& \left(\frac{(H_1-1)\tH_2}{H_{12}} dt \wedge dy_2 \wedge dy_3
 + \frac{(H_2-1)\tH_1}{H_{12}} dt \wedge dy_1 \wedge dy_4 \right) \sin\t\nn
&& \hs{-10} + \frac{1-H_{12}}{H_{12}} dt \wedge dy_1 \wedge dy_3 \cos\t
 + \shalf\frac{(1-H_1)(1-H_2)}{H_{12}} dt \wedge dy_2 \wedge dy_4
 \sin^2 \t \cos\t,
\label{tilt1}
\ena
where we have defined $H_{12}$ by
\bea
H_{12} = \frac{\tH_1 \tH_2 - \cos^2 \t}{\sin^2 \t}.
\ena
{}For generic $\t$, this describes two $2_M$-branes, with one lying along $y_3$
and the direction making an angle $\frac{\pi}{2}+\t$ with $y_1$ and another
along $y_1$ and the direction making an angle $\frac{\pi}{2}+\t$ with $y_3$.
{}For $\t=0$, $H_{12}=H_1+H_2-1$ is a harmonic function and $\tH_1=\tH_2=1$,
and we have a $2_M$-brane with 1/2 supersymmetry. For $\t=\frac{\pi}{2}$,
$H_{12}=H_1 H_2, \tH_1=H_1,\tH_2=H_2$, and we have a $2_M\perp 2_M$, which
has 1/4 supersymmetry. The reason why the number of remaining supersymmetry
changes is that we adopted the convention that the charge is sent to zero
in the infinite boost limit (see (\ref{boost})).

The second possibility is to make boost in $y_4$ direction to produce a bound
state with wave. Then making reduction in $y_4$, T-duality twice along $y_1$
and $x_1$, we get a slightly different tilted $(2_D+2_D\perp 2_D)$ solution,
which can be lifted to 11 dimensions:
\bea
&&ds_{11}^2 = (H_1 \tH_2)^{1/3} \left[ - (H_1 \tH_2)^{-1} dt^2
 + (H_1 \tH_1)^{-1} \left(\tH_1 dy_1 + (H_1-1) \sin\t \cos\t dy_2\right)^2
 \right. \nn
&& \hs{5}\left. + \tH_1^{-1} dy_2^2 + \tH_2^{-1} (dy_3^2 + dy_4^2)
 + H_1^{-1} dx_1^2 + \sum_{i=2}^6 dx_i^2 \right], \nn
&&C = \frac{1-H_1}{H_1} dt \wedge dx_1 \wedge (-dy_1 \cos\t + dy_2 \sin\t)
 + \frac{1-H_2}{\tH_2} \sin^2 \t dt \wedge dy_3 \wedge dy_4.
\label{tilt2}
\ena
Compared with the above boost, this is a solution in which one of the
$2_M$ brane is not tilted.

In this solution, instead of replacing $H_2$ by $\tH_2$, we could have
also kept $H_2$. The resulting solution is then tilted $(2_D\perp 2_D)$
solution obtained by replacing $\tH_2$ in (\ref{tilt2}) by $H_2$.

\subsection{General features of the solutions}

We have seen the effect of boost is just to produce bound
states of wave and the original solutions. This feature is valid not only
in 11 dimensions but also in 10 dimensions. This means that T-duality
after boost is equivalent to making T-duality to bound states with waves.
Then making reductions and T-dualities yields solutions in which branes
are tilted.

These solutions can be understood from the fact that the boost generally
introduces rotations to objects lying in transverse directions with
respect to it. In fact, we have made boost in the direction transverse
to both the $2_M$-branes in the solution (\ref{tilt1}), and we get both
branes tilted. In the second solution (\ref{tilt2}), we have made it
along one of the $2_M$-brane, and that $2_M$-brane in the resulting
solution is not tilted whereas the other is.

We can continue to make similar boost, reduction and dualities with other
configurations such as $2_M\perp 5_M$ or $5_M\perp 5_M$. This has been
discussed in ref.~\cite{BC} for $2_M \perp 5_M$ and in ref.~\cite{CC} for
$(4_D \perp 4_D)_A$ (directly connected to $5_M \perp 5_M$).
The procedure described in the previous section is general, and similar
tilted solutions can be obtained from more general orthogonal intersections
of M-branes like $2_M\perp 2_M \perp 2_M$ by including boost (or wave).
In all these cases, again the basic structure of the resulting solutions
are the same as the solutions (\ref{tilt1}) and (\ref{tilt2}) discussed
in the previous subsection, with angles between the branes, and how they are
tilted are determined by which boost (or wave) one introduces.

Though we did not continue T-dualities on solutions with waves in sect.~3,
we expect that it produces further tilted solutions.

\sect{Construction Rules}

In this section, we discuss the construction rules for 11-dimensional
solutions. From all the examples we have discussed in this paper, we find
that the following rules for the non-marginal non-tilted solutions are
valid (the rules for tilted ones are not discused here):
\begin{enumerate}
\item
To each fundamental $p_i$-brane solution, we assign a harmonic function
$H_i$ depending on the transverse coordinates, and multiply its inverse
to the metric of the coordinates belonging to the $p_i$-brane in the
conformal frame in which the transverse part $\sum_i dx_i^2$ is free.
The rules for other solutions of wave and monopole are similar.
\item
$p$-brane can intersect orthogonally with $p$-brane over $(p-2)$-brane:
wave can intersect with others over a string:
A 2-brane can intersect orthogonally with 5-brane over string, and with
monopole over 2-brane or 0-brane:
A 5-brane can intersect orthogonally with monopole over 5-brane or 3-brane:
A monopole can intersect orthogonally with monopole over 4-brane.
These rules can be read off from the solutions we discussed and are also
given in ref.~\cite{BR}.
\item
To each non-marginal solution with one charge we discussed in sect.~2, we
assign a modified harmonic function $\tH_i$ defined in (\ref{ht}). If the
solution is among intersecting constituents, we multiply its inverse to the
metric of the relevant coordinates such that putting the solutions to the
fundamental ones reduces the configuration compatible with the above rules.
To construct further bound states, we use further modified one (\ref{ht12}).
\item
{}For the non-marginal solutions with two charges, the rules
are as follows: Let us call $0_w \to 2_M \to 5_M \to 0_m$ proper order. Suppose
that the solution is of the type $A_1 \perp B_1 + A_2 \perp B_2$. If the orders
of the solutions $(A_1,A_2)$ and $(B_1,B_2)$ are both in the same order, proper
or its opposite, we use the combination $\tH_{12}$ defined as in (\ref{h12});
if the two orders are opposite, we use the combination $\hat H_{12}$ defined in
(\ref{hh12}). The solutions must be consistent with the orthogonal rules if
one puts the solutions to the orthogonal ones, and also should agree with
the bound states discussed in sect.~2 if one puts one of the charges to zero.
\end{enumerate}

Rules 1 and 2 are the ordinary orthogonal intersection rules. Rule 3 is a
generalization of that given in ref.~\cite{C} for intersections involving the
non-marginal one-charge bound state $(2_M+5_M)$. Though rules 3 and 4 appear
to be ambiguous, the rules are actually useful enough to derive the solutions
we discussed in sects.~2 and 3. We now illustrate how to use these rules
by several examples.

Consider the solution $2_M \perp 2_M + 5_M \perp 5_M$. For $\t=0$,
we know from the orthogonal intersection rules that we should assign the
coordinates to each brane as follows:
\bea
\begin{array}{ccc}
2_M & \perp & 2_M \vs{-3} \\
{\scriptstyle (y_1,y_2)} & & {\scriptstyle (y_3,y_4)} \hs{2}.
\end{array}
\ena
{}For $\t=\frac{\pi}{2}$, this must be $5_M \perp 5_M$ with coordinates
\bea
\begin{array}{ccc}
5_M & \perp & 5_M \vs{-3} \\
{\scriptstyle (y_1,y_2,y_5,y_6,y_7)} & & {\scriptstyle (y_3,y_4,y_5,y_6,y_7)}
 \hs{2},
\end{array}
\hs{5} {\rm or} \hs{5}
\begin{array}{ccc}
5_M & \perp & 5_M \vs{-3} \\
{\scriptstyle (y_1,y_3,y_5,y_6,y_7)} & & {\scriptstyle (y_2,y_4,y_5,y_6,y_7)}
 \hs{2},
\end{array}
\ena
intersecting over 3-brane. These are the only possible configurations to
make the number of world-volume coordinates minimum. For the first case,
the metric for each coordinates must change as\footnote{Here and below,
we discuss metrics up to an overall conformal factor because it is easily
determined by a similar reasoning.}
\bea
&& H_1^{-1}(dy_1^2 + dy_2^2) + H_2^{-1} (dy_3^2 + dy_4^2)
 + dy_5^2 + dy_6^2 + dy_7^2 \nn
\to && H_1^{-1}(dy_1^2 + dy_2^2) + H_2^{-1} (dy_3^2 + dy_4^2)
 + (H_1 H_2)^{-1}(dy_5^2 + dy_6^2 + dy_7^2).
\ena
According to the rule 4 above, we can use $\tH_{12}$ to reproduce this
metric and this gives the solution given in (\ref{s2255}). The second
possibility is excluded because putting $H_1=1$ reduces the solution to
$(2_M + 5_M)$ state but incompatible with (\ref{s25}). Our rule does
forbid this because we cannot reproduce the necessary metric change
by $\tH_{12}$.

Another example is $2_M \perp 5_M + 5_M \perp 2_M$. Let us assign
the coordinates as
\bea
\begin{array}{ccc}
2_M & \perp & 5_M \vs{-3} \\
{\scriptstyle (y_1,y_2)} & & {\scriptstyle (y_1,y_3,\cdots,y_6)}
\end{array}
\to
\begin{array}{ccc}
5_M & \perp & 2_M \vs{-3} \\
{\scriptstyle (y_1,y_2,y_4,y_5,y_6)} & & {\scriptstyle (y_1,y_3)} \hs{2}.
\end{array}
\ena
According to the rule 4, we should use $\hat H_{12}$ to reproduce
the appropriate metric change
\bea
&& (H_1 H_2)^{-1} dy_1^2 + H_1^{-1} dy_2^2 + H_2^{-1} (dy_3^2 +\cdots + dy_6^2)
 \nn
\to && (H_1 H_2)^{-1} dy_1^2 + H_1^{-1} dy_2^2 + H_2^{-1} dy_3^2
 + H_1^{-1} (dy_4^2 +\cdots + dy_6^2),
\ena
and we are lead to the solution given in (\ref{s2552}).

A more subtle case is the solution $2_M \perp 2_M + 5_M \perp 0_w$.
In this case, we can consider two possible choices for the coordinates:
\bea
\begin{array}{ccc}
2_M & \perp & 2_M \vs{-3} \\
{\scriptstyle (y_1,y_2)} & & {\scriptstyle (y_3,y_4)}
\end{array}
\to
\begin{array}{ccc}
5_M & \perp & 0_w \vs{-3} \\
{\scriptstyle (y_1,\cdots,y_5)} & & {\scriptstyle (y_5)} \hs{2},
\end{array}
\hs{4} {\rm or} \hs{4}
\begin{array}{ccc}
5_M & \perp & 0_w \vs{-3} \\
{\scriptstyle (y_1,\cdots,y_5)} & & {\scriptstyle (y_4)} \hs{2}.
\end{array}
\ena
The first case is (\ref{s2250}), but we have not encountered the second.
For the first one, the metric should change as
\bea
&& -(H_1 H_2)^{-1} dt^2 + H_1^{-1} (dy_1^2 + dy_2^2)
 + H_2^{-1} (dy_3^2 + dy_4^2) + dy_5^2 \nn
\to && H_1^{-1}\left[- dt^2 + dy_1^2 +\cdots + dy_5^2
 + (H_2 - 1) (dt + dy_5)^2 \right].
\ena
According to our rule 4, we should use $\hat H_{12}$ for these solutions.
However, the metric for the time coordinate can be reproduced only if we use
$\tH_1$ as well in the numerator as $(H_1{\hat H}_{12})^{-1}\tH_1$.
Thus the precise rule seem to be that only the inverse of the functions
listed in rule 4 should be multiplied to the metric, as in rule 1. By this
rule we precisely obtain the solution (\ref{s2250}). Moreover, the second
possibility is excluded by considering the metric for the coordinate $y_5$,
since its metric changes as $1 \to H_1^{-1}$ which can be reproduced only if
one uses tilde type function in the denominator, in contradiction to the rule.
Again such a solution is excluded by the consistency with the $(0_w+2_M)$
bound state in (\ref{s02}) for $H_1=1$.

Examples of the rule 3 above are (\ref{s2225}), (\ref{s2555}), (\ref{s2022}),
(\ref{s2050}), (\ref{s5052}), (\ref{s2520}), (\ref{s2m5m}), (\ref{s555m})
and (\ref{s5mmm}), and these solutions can be easily reproduced by the above
rules. Other solutions are examples of the rule 4 above.

The generalization of these rules to solutions with more charges is
straightforward. For example, it is easy to make $A_1\perp B_1\perp C_1+
A_2\perp B_2\perp C_2+A_3\perp B_3\perp C_3$ type of bound states by an
obvious generalization of these rules. One can also obtain orthogonal
intersections of these non-marginal and marginal solutions by similar rules.

\sect{Discussions}

We have examined various solutions produced by using the solution
generating technique of reduction, T-duality and lifting. This produces
in general non-marginal as well as tilted brane solutions. On this basis,
we have also presented construction rules for these non-marginal solutions.
We have not included S-duality in these discussions. Let us now discuss
what happens if the S-duality is also taken into account.

It is easy to see how the S-duality transforms 10-dimensional solutions
into others by using the rules given in \cite{BHO}. For type IIB solutions,
this is almost trivial. We find
\bea
1_D \leftrightarrow 1_F, \hs{3}
5_D \leftrightarrow 5_S,
\ena
and other solutions $3_D, 7_D, 0_m$ and wave are invariant.
{}For type IIA solutions, we find
\bea
1_F \left\{
\begin{array}{cl}
\stackrel{y}{\to} & 1_F \\
\stackrel{x}{\to} & 2_D \hs{2},
\end{array} \right.
\hs{5}
2_D \left\{
\begin{array}{cl}
\stackrel{y}{\to} & 1_F \\
\stackrel{x}{\to} & 2_D \hs{2},
\end{array} \right.
\hs{5}
4_D \left\{
\begin{array}{cl}
\stackrel{y}{\to} & 4_D \\
\stackrel{x}{\to} & 5_S \hs{2},
\end{array} \right.
\hs{5}
5_S \left\{
\begin{array}{cl}
\stackrel{y}{\to} & 4_D \\
\stackrel{x}{\to} & 5_S \hs{2},
\end{array} \right. \nn
0_w \left\{
\begin{array}{cl}
\stackrel{y}{\to} & 0_D \\
\stackrel{x}{\to} & 0_w \hs{2},
\end{array} \right.
\hs{5}
0_m \left\{
\begin{array}{cl}
\stackrel{y}{\to} & 0_m \\
\stackrel{x}{\to} & 6_D \hs{2},
\end{array} \right.
\hs{5}
6_D \left\{
\begin{array}{cl}
\stackrel{y}{\to} & 0_m \\
\stackrel{x}{\to} & 6_D \hs{2},
\end{array} \right.
\hs{5}
0_D \left\{
\begin{array}{cl}
\stackrel{x}{\to} & 0_w \hs{2},
\end{array} \right.
\ena
where $y(x)$ stands for world-volume (transverse) direction.
We thus see that the effect of S-duality is the same as the reduction and
lifting discussed in the introduction (\ref{re2M})--(\ref{rem})\footnote{
{}From the 11-dimensional perspective, S-duality can be understood as the
interchange of the 10-th and 11-th coordinates~\cite{BHO}, and thus this
is a simple consequence of this fact.} and hence at least part of the effect
of S-duality is incorporated if we find solutions in 11-dimensional theory.

Let us discuss the general feature of the ADM mass formula for these non-tilted
non-marginal solutions with unbroken supersymmetries. It is not difficult
to show that the non-marginal solutions with one charge typically have the mass
formula with single term like
\bea
m \sim \sqrt{Q_1^2 + Q_2^2 + \cdots},
\ena
where the number of charges depends on how many electric and magnetic
charges are involved in the solution. For example, the three-body solution
(\ref{sw25}) has
\bea
Q_e = Q \sin\t_1 \cos\t_2, \hs{3}
Q_m = Q \sin\t_1 \sin\t_2, \hs{3}
Q_3 = Q \cos\t_1,
\ena
and 
\bea
m \sim \sqrt{Q_e^2 + Q_m^2 + Q_3^2}.
\ena

{}For the non-marginal solutions with two charges, the formula typically
takes the sum of two terms
\bea
m \sim \sqrt{Q_1^2 + Q_2^2 + \cdots} + \sqrt{{Q_1'}^2 + {Q_2'}^2 + \cdots},\ena
{}For example, we get from (\ref{s2255})
\bea
Q_{e1} = Q_1 \cos\t, \hs{3}
Q_{e2} = Q_2 \cos\t, \hs{3}
Q_{m1} = Q_1 \sin\t, \hs{3}
Q_{m2} = Q_2 \sin\t,
\ena
so that the mass is given by
\bea
m \sim \sqrt{Q_{e1}^2 + Q_{m1}^2} + \sqrt{Q_{e2}^2 + Q_{m2}^2}.
\ena

We can continue this analysis for three-charge solutions and so on.
Given such a large number of classical non-marginal solutions with unbroken
supersymmetries, it would be interesting to examine their quantum properties.
In this connection, we point out that all the non-marginal solutions found
in this paper have an interesting property. If we calculate the determinant
of the metrics for the nine-space part, which are relevant to the nine-area
calculation, we find that tilde or hatted functions all cancel out from
the expression for non-marginal solutions discussed in sects.~2 and 3.
This means that the nine-area or entropy for these non-marginal solutions
are independent of the angles which characterize the non-marginality.
This is equivalent to T-duality invariance,\footnote{S-duality is manifest in
11 dimensions, so that the entropy is U-invariant.} and we expect
the same property is valid for other non-marginal solutions involving more
than two charges.

\section*{Acknowledgement}
This work was supported in part by the Japan Society for the Promotion of
Science, and in part by Grant-in-aid from the Ministry of Education, Science,
Sports and Culture.

\appendix

\sect{Appendix: Reduction, T-duality and lifting rules}

In this appendix, we summarize the reduction, T-duality and lifting
rules~\cite{BHO} which are heavily used in the text. We use the same
convention as ref.~\cite{BHO}.

\noindent{\bf
11 D SUGRA $\to$ IIA:}

{}From 11-dimensional supergravity with the metric $\hat{\hat{g}}_{\hat{\mu}
\hat{\nu}}$ and a 3-form $\hat{C}_{\hat{\mu}\hat{\nu}\hat{\rho}}$,
upon dimensional reduction in ${\underline y}$, we get type IIA supergravity
with a metric, 3-, 2- and 1-forms and dilaton:
\bea
\hat{g}_{\hat{\mu}\hat{\nu}}
&=& \left( \hat{\hat{g}}_{\underline{y}\underline{y}} \right)^{\frac{1}{2}}
 \left (\hat{\hat{g}}_{\hat{\mu}\hat{\nu}}
 -\hat{\hat{g}}_{\hat{\mu}\underline{y}}
 \hat{\hat{g}}_{\hat{\nu}\underline{y}}
 /\hat{\hat{g}}_{\underline{y}\underline{y}} \right)\, ,\hs{10}
\hat{C}_{\hat{\mu}\hat{\nu}\hat{\rho}}
= \hat{\hat{C}}_{\hat{\mu}\hat{\nu}\hat{\rho}}\, , \nn
\hat{A}^{(1)}_{\hat{\mu}} &=& \hat{\hat{g}}_{\hat{\mu}\underline{y}}
 /\hat{\hat{g}}_{\underline{y}\underline{y}}\, , \hs{45}
\hat{B}^{(1)}_{\hat{\mu}\hat{\nu}} = 
 \frac{3}{2}\hat{\hat{C}}_{\hat{\mu}\hat{\nu}\underline{y}}\, , \nn
\hat{\phi} &=& 
 \frac{3}{4}\log{\left( \hat{\hat{g}}_{\underline{y}\underline{y}}\right)}\, .
\label{10ddef}
\ena

\noindent{\bf
IIA $\to$ 11 D SUGRA:}

Conversely 11-dimensional supergravity is recovered by the formula
\bea
\hat{\hat{g}}_{\hat{\mu}\hat{\nu}}
&=& e^{-\frac{2}{3}\hat{\phi}}\hat{g}_{\hat{\mu}\hat{\nu}}
 +e^{\frac{4}{3}\hat{\phi}}\hat{A}^{(1)}_{\hat{\mu}}
 \hat{A}^{(1)}_{\hat{\nu}}\, ,\hspace{1cm}
\hat{\hat{C}}_{\hat{\mu}\hat{\nu}\hat{\rho}}
= \hat{C}_{\hat{\mu}\hat{\nu}\hat{\rho}}\, , \nn
\hat{\hat{g}}_{\hat{\mu}\underline{y}}
&=& e^{\frac{4}{3}\hat{\phi}}\hat{A}^{(1)}_{\hat{\mu}}\, , \hs{36}
\hat{\hat{C}}_{\hat{\mu}\hat{\nu}\underline{y}}
= \frac{2}{3}\hat{B}^{(1)}_{\hat{\mu}\hat{\nu}}\, , \nn
\hat{\hat{g}}_{\underline{y}\underline{y}} &=&
e^{\frac{4}{3}\hat{\phi}}\, .
\ena

\noindent{\bf
T-duality (type--IIB $\to$ type--IIA):}

The T-duality rules from type--IIB to type--IIA are
\bea
\hat{g}_{\mu\nu}
&=& \hat{\jmath}_{\mu\nu}
 -\left( \hat{\jmath}_{\underline{x}\mu}\hat{\jmath}_{\underline{x}\nu}
 -\hat{\cal B}^{(1)}_{\underline{x}\mu}
 \hat{\cal B}^{(1)}_{\underline{x}\nu} \right)
 /\hat{\jmath}_{\underline{x}\underline{x}}\, , \nn
\hat{g}_{\underline{x}\mu}
&=& \hat{\cal B}_{\underline{x}\mu}^{(1)}
 /\hat{\jmath}_{\underline{x}\underline{x}}\, , \hs{10}
\hat{g}_{\underline{x}\underline{x}}
= 1/\hat{\jmath}_{\underline{x}\underline{x}}\, , \nn
\hat{C}_{\underline{x}\mu\nu}
&=& {\textstyle\frac{2}{3}}\left[ \hat{\cal B}_{\mu\nu}^{(2)}
 +2 \hat{\cal B}^{(2)}_{\underline{x}[\mu}\hat{\jmath}_{\nu]\underline{x}}
 /\hat{\jmath}_{\underline{x}\underline{x}} \right]\, , \nn
\hat{C}_{\mu\nu\rho}
& = & {\textstyle\frac{8}{3}} \hat{D}_{\underline{x}\mu\nu\rho}
 +\epsilon^{ij}\hat{\cal B}^{(i)}_{\underline{x}[\mu}
 \hat{\cal B}^{(j)}_{\nu\rho]}
 +\epsilon^{ij}\hat{\cal B}^{(i)}_{\underline{x}[\mu}
 \hat{\cal B}^{(j)}_{|\underline{x} |\nu} \hat{\jmath}_{\rho]\underline{x}}
 /\hat{\jmath}_{\underline{x}\underline{x}}\, ,\nn
\hat{B}^{(1)}_{\mu\nu}
&=& \hat{\cal B}^{(1)}_{\mu\nu} +2\hat{\cal B}^{(1)}_{\underline{x}[\mu}
 \hat{\jmath}_{\nu]\underline{x}}
 /\hat{\jmath}_{\underline{x}\underline{x}}\, ,\hs{10}
\hat{B}_{\underline{x}\mu}^{(1)}
= \hat{\jmath}_{\underline{x}\mu}
 /\hat{\jmath}_{\underline{x}\underline{x}}, \nn
\hat{A}^{(1)}_{\mu}
&=& -\hat{\cal B}_{\underline{x}\mu}^{(2)} +\hat{\ell}
 \hat{\cal B}_{\underline{x}\mu}^{(1)}\, , \hs{10}
\hat{A}_{\underline{x}}^{(1)} = \hat{\ell}\, , \nn
\hat{\phi}
&=& \hat{\varphi} -\frac{1}{2}
 \log{(\hat{\jmath}_{\underline{x}\underline{x}})}\, .
\label{TBA}
\ena

\noindent{\bf
T-duality (type--IIA $\to$ type--IIB):}

The converse rules are
\bea
\hat{\jmath}_{\mu\nu}
&=& \hat{g}_{\mu\nu} -\left(\hat{g}_{\underline{x}\mu}
 \hat{g}_{\underline{x}\nu}
 -\hat{B}^{(1)}_{\underline{x}\mu}
 \hat{B}^{(1)}_{\underline{x}\nu}\right)
 /\hat{g}_{\underline{x}\underline{x}}\, , \nn
\hat{\jmath}_{\underline{x}\mu}
&=& \hat{B}^{(1)}_{\underline{x}\mu}
 /\hat{g}_{\underline{x}\underline{x}}\, , \hs{10}
\hat{\jmath}_{\underline{x}\underline{x}}
= 1/\hat{g}_{\underline{x}\underline{x}}\, , \nn
\hat{D}_{\underline{x}\mu\nu\rho}
&=& {\textstyle\frac{3}{8}}\left[
 \hat{C}_{\mu\nu\rho} -\hat{A}^{(1)}_{[\mu}\hat{B}^{(1)}_{\nu\rho]}
 +\hat{g}_{\underline{x}[\mu}\hat{B}^{(1)}_{\nu\rho]}
 \hat{A}^{(1)}_{\underline{x}}/\hat{g}_{\underline{x}\underline{x}}
 -{\textstyle\frac{3}{2}}\hat{g}_{\underline{x}[\mu}
 \hat{C}_{\nu\rho]\underline{x}}
 /\hat{g}_{\underline{x}\underline{x}}\right]\, ,\nn
\hat{\cal B}^{(1)}_{\mu\nu}
&=& \hat{B}^{(1)}_{\mu\nu} +2\hat{g}_{\underline{x}[\mu}
 \hat{B}^{(1)}_{\nu]\underline{x}}
 /\hat{g}_{\underline{x}\underline{x}}\, , \hs{10}
\hat{\cal B}^{(1)}_{\underline{x}\mu}
= \hat{g}_{\underline{x}\mu}/\hat{g}_{\underline{x}\underline{x}}\, ,\nn
\hat{\cal B}^{(2)}_{\mu\nu}
&=& {\textstyle\frac{3}{2}}\hat{C}_{\mu\nu\underline{x}}
 -2 \hat{A}^{(1)}_{[\mu}\hat{B}^{(1)}_{\nu]\underline{x}}
 +2\hat{g}_{\underline{x}[\mu}\hat{B}^{(1)}_{\nu]\underline{x}}
 \hat{A}^{(1)}_{\underline{x}}/\hat{g}_{\underline{x}\underline{x}}\, , \nn
\hat{\cal B}^{(2)}_{\underline{x}\mu}
&=& -\hat{A}^{(1)}_{\mu} +\hat{A}^{(1)}_{\underline{x}}
 \hat{g}_{\underline{x} \mu}/\hat{g}_{\underline{x}\underline{x}}\, , \nn
\hat{\varphi} &=& \hat{\phi}-\frac{1}{2}\log{(\hat{g}_{\underline{x}
\underline{x}})}\, , \hs{10}
\hat{\ell} = \hat{A}_{\underline{x}}^{(1)}\, .
\label{TAB}
\ena

\newcommand{\NP}[1]{Nucl.\ Phys.\ {\bf #1}}
\newcommand{\AP}[1]{Ann.\ Phys.\ {\bf #1}}
\newcommand{\PL}[1]{Phys.\ Lett.\ {\bf #1}}
\newcommand{\NC}[1]{Nuovo Cimento {\bf #1}}
\newcommand{\CMP}[1]{Comm.\ Math.\ Phys.\ {\bf #1}}
\newcommand{\PR}[1]{Phys.\ Rev.\ {\bf #1}}
\newcommand{\PRL}[1]{Phys.\ Rev.\ Lett.\ {\bf #1}}
\newcommand{\PRE}[1]{Phys.\ Rep.\ {\bf #1}}
\newcommand{\PTP}[1]{Prog.\ Theor.\ Phys.\ {\bf #1}}
\newcommand{\PTPS}[1]{Prog.\ Theor.\ Phys.\ Suppl.\ {\bf #1}}
\newcommand{\MPL}[1]{Mod.\ Phys.\ Lett.\ {\bf #1}}
\newcommand{\IJMP}[1]{Int.\ Jour.\ Mod.\ Phys.\ {\bf #1}}
\newcommand{\JP}[1]{Jour.\ Phys.\ {\bf #1}}

\end{document}